\documentclass[conference]{IEEEtran}
\IEEEoverridecommandlockouts
\usepackage{cite}
\usepackage[utf8]{inputenc}
\usepackage{graphicx}
\usepackage{amsmath}
\usepackage[version=4]{mhchem}
\usepackage{siunitx}
\usepackage{longtable,tabularx}
\usepackage{subcaption}
\usepackage{epstopdf}
\usepackage{tabularx}
\usepackage{booktabs} 
\usepackage{multirow}
\usepackage{arydshln} 
\usepackage{url}
\usepackage{float}
\usepackage{acronym}
\usepackage{fancyhdr}
\fancypagestyle{specialfooter}{%
  \fancyhf{}
  
  \fancyfoot[L]{\footnotesize This work was presented at the AIAA Aviation 2023 Forum. \\ Copyright @ Souma Chowdhury}
  \fancyfoot[C]{1}
}
\usepackage{hyperref} 
\usepackage{textcomp} 
\def\BibTeX{{\rm B\kern-.05em{\sc i\kern-.025em b}\kern-.08em
    T\kern-.1667em\lower.7ex\hbox{E}\kern-.125emX}}

\renewcommand{\footnoterule}{%
  \kern -3pt
  \hrule width \columnwidth height 1pt
  \kern 2pt
}

\begin{document}

\title{Learning Constrained Corner Node Trajectories of a Tether Net System for Space Debris Capture}

\makeatletter
\newcommand{\linebreakand}{%
  \end{@IEEEauthorhalign}
  \hfill\mbox{}\par
  \mbox{}\hfill\begin{@IEEEauthorhalign}
}
\makeatother

\author{\IEEEauthorblockN{Feng Liu$^*$ \thanks{$^*$ Ph.D. Student, Department of Mechanical and Aerospace Engineering, AIAA Student Member}}
\IEEEauthorblockA{\textit{University at Buffalo}\\
Buffalo, New York, 14260 \\
fliu23@buffalo.edu}
\and
\IEEEauthorblockN{Achira Boonrath$^*$}
\IEEEauthorblockA{\textit{University at Buffalo}\\
Buffalo, New York, 14260 \\
achirabo@buffalo.edu}
\and
\IEEEauthorblockN{Prajit KrisshnaKumar$^*$}
\IEEEauthorblockA{\textit{University at Buffalo}\\
Buffalo, New York, 14260 \\
prajitkr@buffalo.edu}
\linebreakand
\IEEEauthorblockN{Eleonora M. Botta$^\dagger$ \thanks{$^\dagger$ Assistant Professor, Mechanical and Aerospace Engineering, AIAA member}}
\IEEEauthorblockA{\textit{University at Buffalo}\\
Buffalo, New York, 14260 \\
ebotta@buffalo.edu}
\and
\IEEEauthorblockN{Souma Chowdhury$^\ddag$ \thanks{$^\ddag$ Associate Professor, Mechanical and Aerospace Engineering, AIAA Senior Member, Corr. author}}
\IEEEauthorblockA{\textit{University at Buffalo}\\
Buffalo, New York, 14260 \\
soumacho@buffalo.edu}
}

\maketitle
\thispagestyle{plain}
\pagestyle{plain}

\thispagestyle{specialfooter}

\begin{abstract}
The earth’s orbit is becoming increasingly crowded with debris that poses significant safety risks to the operation of existing and new spacecraft and satellites. The active tether-net system, which consists of a flexible net with maneuverable corner nodes launched from a small autonomous spacecraft, is a promising solution for capturing and disposing of such space debris. The requirement of autonomous operation and the need to generalize over scenarios with debris scenarios in different rotational rates makes the capture process significantly challenging. The space debris could rotate about multiple axes, which, along with sensing/estimation and actuation uncertainties, calls for a robust, generalizable approach to guiding the net launch and flight – one that can guarantee robust capture. This paper proposes a decentralized actuation system combined with reinforcement learning for planning and controlling this tether-net system. In this new system, four microsatellites with cold gas type thrusters act as the corner nodes of the net and can thus help control or correct the flight of the net after launch. The microsatellites pull the net to complete the task of approaching and capturing the space debris. The proposed method uses a RL framework that integrates a proximal policy optimization to find the optimal solution based on the dynamics simulation of the net and the microsatellites performed in Vortex Studio. The RL framework finds the optimal trajectory that is both fuel-efficient and ensures a desired level of capture quality.
\end{abstract}

\newacro{ADR}{Active Debris Removal}
\newacro{RL}{Reinforcement Learning}

\begin{IEEEkeywords}
Active debris removal, reinforcement learning, optimization
\end{IEEEkeywords}


\section{Introduction}
Earth’s orbit is becoming increasingly dangerous for current and future space missions since the growing amount of space debris threatens operational safety \cite{danger}. \ac{ADR} is one of the solutions to mitigate the problem. Among the multiple methods studied, tether-net systems have been proposed for their high flexibility and good capturing range \cite{flex}. Previous studies \cite{net1,net2,net3} have shown that tether-net systems are effective for capturing uncooperative debris. Among others, the research of Botta et al. \cite{bnet1,bnet2,bnet3} examined the dynamics of the deployment and capture phase of the debris removal tasks using net-based systems. Chen et al. analyzed the system's robustness to errors in a sample mission scenario in which the second stage of the Zenit-2 launch vehicle is the target debris of interest \cite{chen2022analysis}. Additionally, Zeng et al. \cite{zeng} conducted research on the closing mechanism with uncertainties, which applies \ac{RL} \cite{rl} to ensure debris capture.

Studies \cite{sp_robot2} have shown that using space robots is effective in increasing the efficiency and reliability of capturing uncooperative space debris. Meng et al. \cite{approach} proposed the Autonomous Maneuverable Space Net (AMSN) system, which consists of a flexible net and several Maneuverable Units (MUs). The AMSN has a greater effective net deployment range than the traditional tether-net systems, and the MUs allow the AMSN to perform more flexible operations. A chaser satellite brings the AMSN to rendezvous with the target in orbit around the Earth. The chaser then releases the AMSN with an initial velocity, and the MUs on the AMSN control the shape and movement of the net. The net closes, and locks after the target is in the net mouth. The AMSN then drags the captured target into the atmosphere to be incinerated or to a graveyard orbit. In this process, the trajectory and shape of the net are essential for a successful mission \cite{approach}.

In intelligent autonomous systems, Artificial Neural Networks (ANN) have become a promising analysis tool for decision-support models \cite{ann}. An ANN can map states to actions in a policy model, and various ANN learning methods have already been applied to robotics and control applications. Besides \ac{RL} \cite{rl}, learning methods such as Neuroevolution \cite{neve} and Supervised Learning \cite{supl} are also popular in similar scenarios. For the tether-net systems, the launching and wrapping control is compatible with the advanced \ac{RL} \cite{arl} and neuroevolution \cite{neruoev} methods. Due to various debris characteristics, the system uncertainties and selecting optimal actions could be challenging without these learning methods.

Most previous studies are based on the assumption that the launching phase is under ideal conditions. However, in reality, the perfect launching conditions are challenging due to uncertainties and hardware limitations for both the launch equipment and the determination of the target pose. The deployment trajectory of the net is one of the most critical components for a successful capture, especially when considering that a relaunch of the system is not possible, so a method to overcome the effect of potential error and discover more reliable trajectories for the MUs to follow needs to be developed.

Meanwhile, most of the designs of tether-net systems are centralized \cite{approach}, which works well in a short deployment distance and in an ideal environment where the uncertainties in the mission are kept to the minimum, and the target is rotating slowly. A semi-decentralized system offers more flexibility and robustness in this complex environment for a long-range deployment, with a target having more complex movement, such as spinning about multiple axes, and environmental uncertainty. Therefore, instead of focusing on the control of the system in a centralized method, focusing on learning the corner nodes' trajectory to design a semi-decentralized system can be a different approach to capturing the target.

The succeeding sections of this paper are arranged as the following: Section \ref{FrameworkoftheModel} describes the machine learning framework utilized in the studies conducted within this paper. Section \ref{DynamicsofDMSNModel} details the dynamics modeling of the maneuverable tether-net system. Section \ref{LearningtheOptimalApproachingandClosing Policies} examines the design optimization approach taken for the maneuverable tether-net system configuration.

\section{Overall Framework for Learning Corner Node Control}\label{FrameworkoftheModel}
The trajectory is one of the key elements for controlling the capture process, and finding the optimal trajectory for the corner nodes can lead to a successful capture. This paper proposes a semi-decentralized reinforcement-learning-based maneuverable space net (RMSN) inspired by the design of Meng et al. \cite{approach}, and the extensive dynamics research of Botta et al. \cite{bnet1,bnet2}. The machine learning framework is inspired by the study of Zeng et al. \cite{zeng}. Compared to non-autonomous tether-net systems, RMSN has a further capture distance and more flexible maneuverability like the AMSN. Meanwhile, RMSN is even more flexible than AMSN due to its semi-decentralized property and is more robust for capturing a target with more complex movement. The machine-learning-based policy optimization of this robotics system makes RMSN more adaptable and robust under uncertainties. The process is split into two phases for the case study: approaching and capturing. The approaching phase starts with the net launching and ends when the net is just about to contact the target debris. The capturing phase follows the approaching phase and ends when the net is closed and the debris is captured. 

The reinforcement learning technique used in this paper is Proximal Policy Optimization (PPO) from stable baselines \cite{stable-baselines}. As a state-of-the-art \ac{RL} method, PPO has proved to be efficient, adaptable, and reliable. By interacting with the environment, PPO updates the gradient based on the experience. Once the update completes, the collected experience is no longer used, so that the next update will start with the new experience. The policy learning framework is showing in Fig. \ref{fig:Network}, which is inspired by the work of Zeng et al. \cite{zeng}. The neural network takes the target's Z-axis offset as input and generates a set of thrust angles as actions to maneuver the net. The final capture quality and fuel consumption are evaluated to calculate rewards for updating the policy.

\begin{figure}[t]
    \centering
    \includegraphics[width=0.5\textwidth]{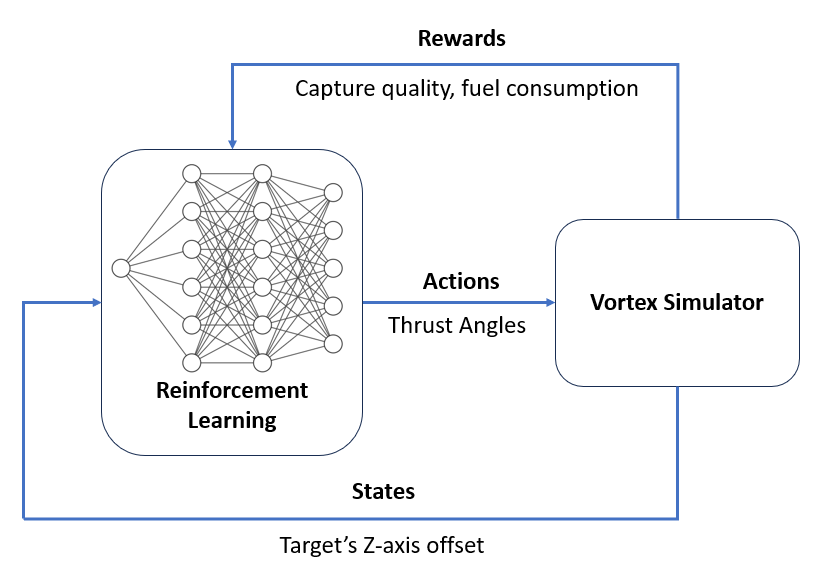}
    \caption{Proposed Policy Learning Process} 
    \label{fig:Network}
\end{figure}

The machine learning framework considers environmental parameters (including geometry and states of MUs, target, and the chaser) and uncertainties of the initial distance between the chaser and the target. The framework finds the optimal policy for each phase to maximize the probability of successful capture, evaluated by the Capture Quality Index (CQI), and minimize fuel consumption. The thrust angles control the corner nodes' trajectories, and by tuning the thrust angles, the optimal trajectory with the highest success rate and minimum fuel cost can be found. Figure \ref{fig:flow} shows the workflow of \ac{RL} of this paper. To make the simulation reflect some realistic problems, such as errors in the sensor readings, the angular velocity of the chaser, and inaccurate launching velocity, noises need to be added to the simulation in the presence of uncertainty. In this paper, the target is set to have a 9 m offset on X-axis, so it is not aligned with the centerline of the net. A noise ranging from -5 m to +5 m was also applied to the Z-axis position of the target. After initialization, the \ac{RL} policy model receives observations from the Vortex Studio-based tether-net simulator and generates actions to be the new input to the simulator. The outputs from the simulator are used to calculate the reward, which is then used to update the policy model.


\begin{figure*}[!t]
    \centering
    \includegraphics[width=1\textwidth]{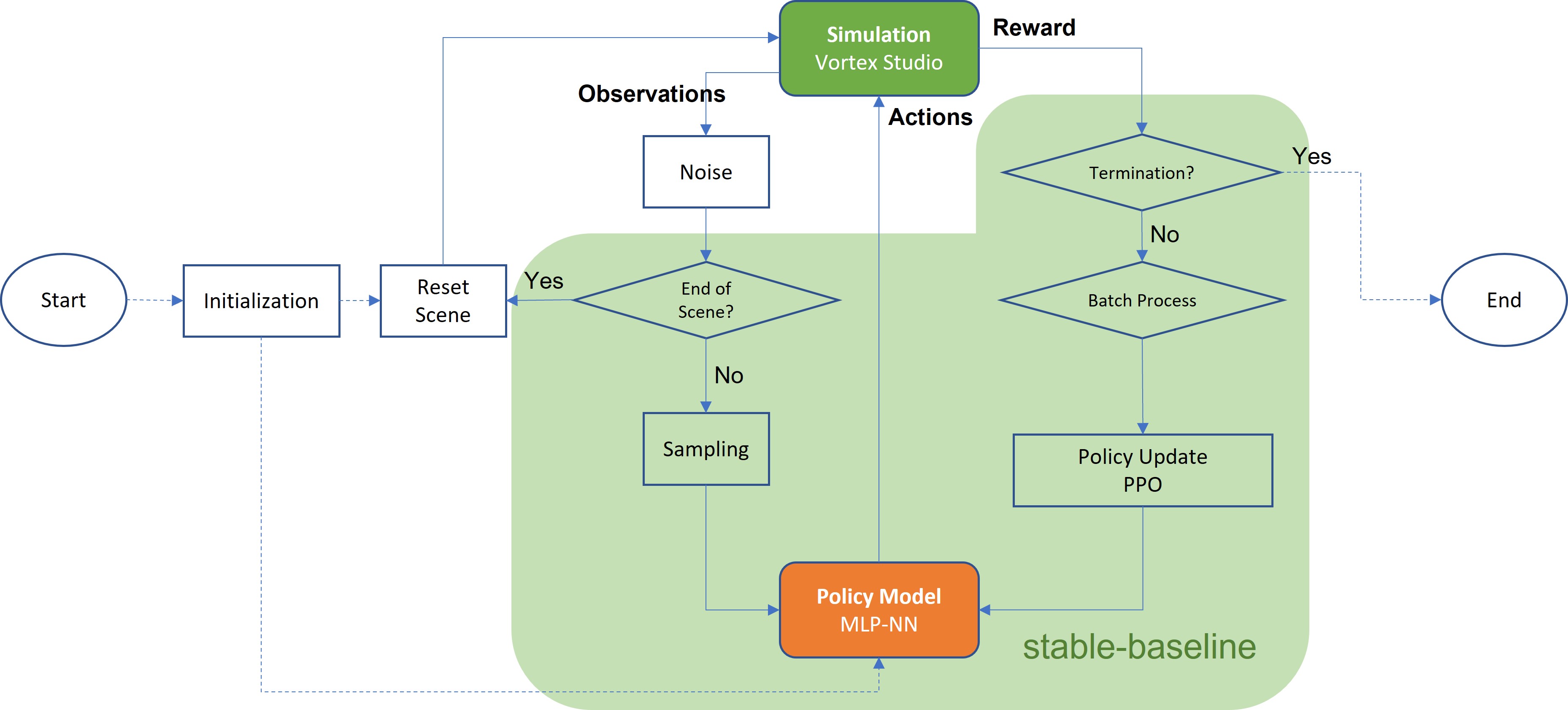}
    \caption{The Workflow of \ac{RL} to Design Corner Node Control for the Tether-Net Capture Mission} 
    \label{fig:flow}
\end{figure*}

\begin{figure}[h]
    \centering
    \includegraphics[width=0.49\textwidth]{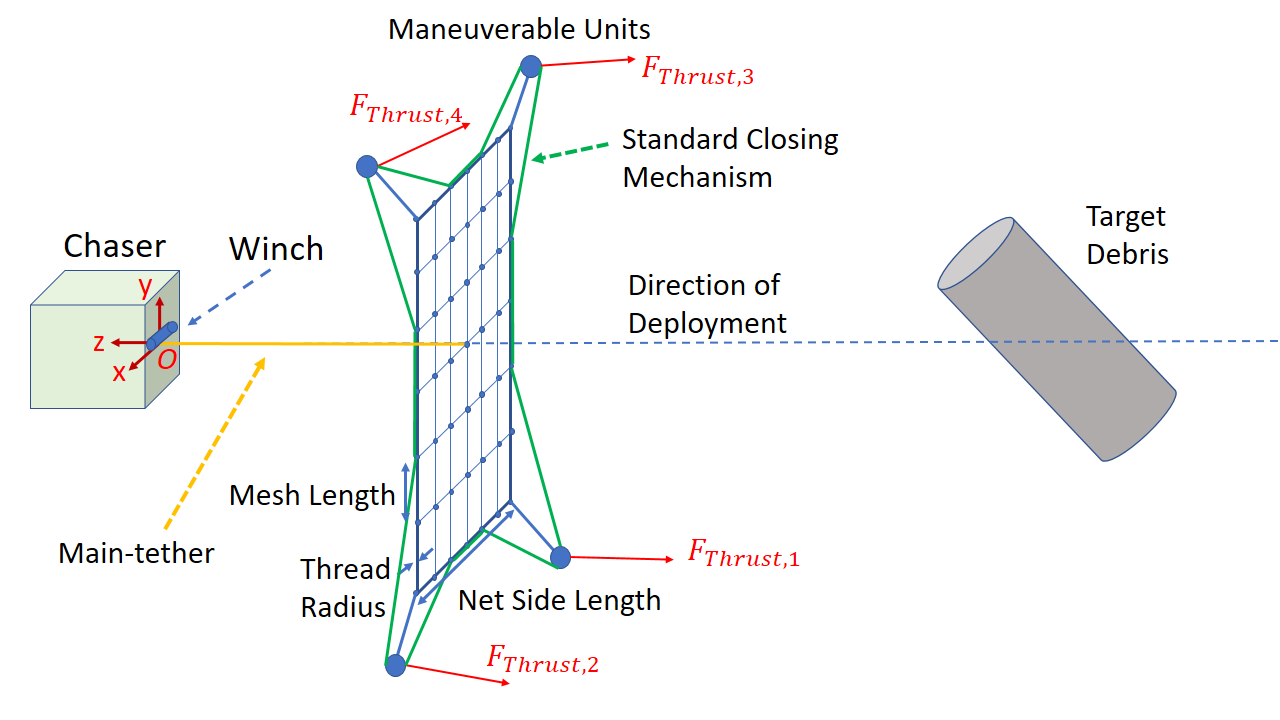}
    \caption{Sketch of the Modeled Tether-net System}
    \label{fig:sket}
\end{figure}

\section{Model of Space Tether-Net }\label{DynamicsofDMSNModel}
The system consists of a square-shaped net, a tether that connects the net to a chaser vehicle, and four MUs. Fig.  \ref{fig:sket} shows the structure of the net, chaser, MUs, main-tether, winch, and closing mechanism. The MUs in this proposal can be understood as miniature satellites \cite{microsat} with thrusters. The simulator used in this paper is based in Vortex Studio, a multi-body dynamics simulation software. Inherited from the work of Botta et al. \cite{bnet3}, the mass of the net is lumped into multiple small spherical rigid bodies at the knots of the net and its corner MU elements, both of which are called \emph{nodes}. The axial stiffness and damping properties of the threads in the net are modeled as springs and dampers in parallel between the nodes that cannot withstand compression.

The mass lumped in the \(j\)-th node, \(m_j\), is defined in the following equation \cite{net3}:
\begin{equation}
    \label{eq:seq}
 m_j = 
\begin{cases}
    \sum_{\gamma \epsilon \Gamma_j} \frac {m_\gamma}{2}+m_{knot}\:& j\:=\:1:N_s^2\\
    \sum _{\gamma	\epsilon \Gamma_j} \frac{m_\gamma}{2}+m_{MU} & j=N^2_s+1\::\:N^2_s+4\\
\end{cases}
\end{equation}
where $m_{\gamma}$ is the mass of the threads adjacent to the \(j\)-th node belonging to set \(\Gamma_j\), \(N_s^2\) is the total number of nodes in the net, \(m_{knot}\) is the mass of the knots of the net where the threads intersect, and \(m_{MU}\) is the total mass of each MU. The equations of motion of the nodes are obtained by writing Newton's second law:

\begin{equation}
    \label{eq:seq1}
        m_j\mathbf{a}_j=\sum_{\gamma\ \epsilon  \Gamma}\pm \mathbf{T}_\gamma + \sum_{s=1}^{S_j}\mathbf{F}_{ext,s,j}
\end{equation}
\noindent where \(\mathbf{a}_j\) is the absolute acceleration of \(j\)-th node; \(\mathbf{T}_\gamma\) is the tension forces in the thread adjacent to the \(j\)-th node; \(\mathbf{F}_{ext,s,j}\) is each of the external forces on the \(j\)-th node. The external forces include forces generated by thrusters, contact forces, and gravitational forces. For the scenarios within this paper, the gravitational acceleration is neglected. The tension force is obtained by writing:
\begin{equation}
    \label{eq:seq2}
\mathbf{T}_\gamma = \left\{\begin{matrix}
T_\gamma \mathbf{e}_\gamma & \textrm{if }(l_\gamma > l_{\gamma,0}) \\
\mathbf{0} & \textrm{if }(l_\gamma \le l_{\gamma,0}) \\
\end{matrix}\right.
\end{equation}

\noindent The magnitude of the tension \(T_\gamma\) can be calculated with \(T_\gamma = k_{a,\gamma}(l_\gamma-l_{\gamma,0}) + c_{a,\gamma}v_{r,\gamma}\). The vector $\mathbf{e}_\gamma$ is axial unit vector of the \(\gamma\)-th thread; \(k_{a,\gamma}\) and \(c_{a,\gamma}\) are stiffness and damping coefficients of the \(\gamma
\)-th thread; \(l_\gamma\) is the current length of the thread; \(l_{\gamma,0}\) is the unstretched length of the thread. $v_{r,\gamma}$ is the projection of the relative velocity of the thread end nodes in the axial direction.

Each rigid body is assigned a material and a collision geometry to model contact dynamics. At each timestep, the simulator checks for the contact between rigid bodies, and contact forces are computed when it is detected. The contact forces rely on the constraint of no penetration between the rigid bodies and the relative velocities of the bodies in contact. The frictional contact forces are calculated using the scaled-box friction model -- an approximation of Coulomb's friction modeling -- while the normal contact forces and contact forces normal to the plane of contact are computed following a modified Kelvin-Voigt model and the Hertzian theory, respectively. Interested readers should reference \cite{net3} for more information regarding contact modeling.

The direction of deployment is the negative Z direction on the coordinates chosen. The net's MUs are given an initial velocity with a magnitude of $v_e$, their components in the X and Y directions have the same magnitude and are defined by the following expressions:
\begin{equation}
    v_{x,\:0}=v_{y,\:0}=v_e \sin{\left(\theta_e \right)}/\sqrt{2}
\end{equation}

\noindent The shooting angle, denoted by $\theta_e$, is defined as the angle between the initial velocity vector of each MU and the direction of deployment. The magnitude of the initial velocity vector in the direction of deployment can be expressed as:  
\begin{equation}
    v_{z,0}=v_e \cos{\left(\theta_e \right)}
\end{equation}

A cubic chaser spacecraft with a side length of $L_{ch} $ and mass $m_{ch}$ is utilized to bring the tether-net system close to the target and to move debris into a disposal orbit. The chaser spacecraft is allowed to float freely without any control in the scenarios considered. The main tether, modeled using multiple slender rigid bodies attached via relaxed prismatic joints that accommodates the simulation of axial, bending, and torsional stiffness, connects the center node of the net to a winch with mass $m_w$ and radius $r_w$. The winch is set to be free to spool during deployment and locked when the closing mechanism is activated and located on one side of the chaser. For the scenarios considered in this paper, as mentioned in \cite{net3}, torsional stiffness is deemed negligible and is therefore not included. The main tether has a density, Young's modulus, cross-sectional radius, and length of $ \rho_t$, $E_t$,  $r_t$, and $L_t$, respectively. The axial stiffness is computed with the following expression per unit length:

\begin{equation}
    EA\:=E_t\pi r^2_t
\end{equation}
Meanwhile, the bending stiffness per unit length $ EI $ is written as: 
\begin{equation}
    EI=\frac{E_t\pi \:r^4_t}{4}
\end{equation}

A set of threads is used for the closing mechanism, which passes through the four MUs and eight nodes on the net's perimeter. In the current design, the closing mechanism is activated by four winches placed in each of the MUs to allow for independence from the main tether. The activation of the closing mechanism is represented by applying a constant force between the attachment points of the closing mechanism until distances between adjacent points become less than a desired distance \cite{bnet3}, chosen to be 2.0 m for the scenarios of interest. Once the desired adjacent length is achieved, a constraint is applied to lock adjacent pairs of attachment points. As such, there can be a maximum of $N_L=12$ locked pairs for the net geometry used in this paper. In future designs, the MU's themselves may be used to close the mouth of the net. However, this will require the development of a complex movement coordination algorithm between the MUs.

Each MU is modeled as a spherical rigid body with radius $r_{MU}$, which is attached to the net proper by corner threads with radius $r_{CT}$ and length $l_{CT}$. To control the MUs, open-loop thrust forces $\mathbf{F}_{Thrust,i}$ are applied. The thrusters are activated at $t=15.0$ s after ejection to allow the net to be sufficiently open and are switched off when the center of mass of the net and the target are a set distance from each other. Each $i$-th thruster is assigned a constant magnitude of $F_{Thrust}=8.9$ N and a constant propellant consumption rate of 0.0121 kg/s based on the cold gas thruster datasheet of VACCO \cite{thrusters}. The components of the thrust in the X, Y, and Z directions are defined as:

\begin{subequations}
\label{eq:thrusts}
\begin{align}
\mathbf{F}_{Thrust_i}\:=F_{Thrust_i,x}\mathbf{\hat{i}}+F_{Thrust_i,y}\mathbf{\hat{j}}+F_{Thrust_i,z}\mathbf{\hat{k}}\\
  F_{Thrust_i,x}=F_{Thrust} \sin(\theta_{Thrust})\cos(\psi_{Thrust_i})\\
F_{Thrust_i,y}=F_{Thrust} \sin(\theta_{Thrust})\sin(\psi_{Thrust_i})\\
F_{Thrust_i,z}=F_{Thrust} \cos(\theta_{Thrust})
\end{align}
\end{subequations}

\noindent where the angles $\psi_{Thrust,i}$ in the X-Y plane and $\theta_{Thrust}$ Z-Y are visualized in the diagram in Fig \ref{fig:Tangles}. Each thrust has a unique angle in the X-Y plane but a common angle in the Z-Y plane. This work assumes that the MUs have the attitude control capability to direct the thrusters in the desired directions throughout their activation. The physical parameters of the system, particularly those for the chaser, main tether, and winch, as well as properties of the net except for the thread radius and initial conditions, are inherited from previous work \cite{tab} and are summarized in Table \ref{tab:net}.

\begin{figure*}[!t]
    \centering
    \includegraphics[width=0.5\textwidth]{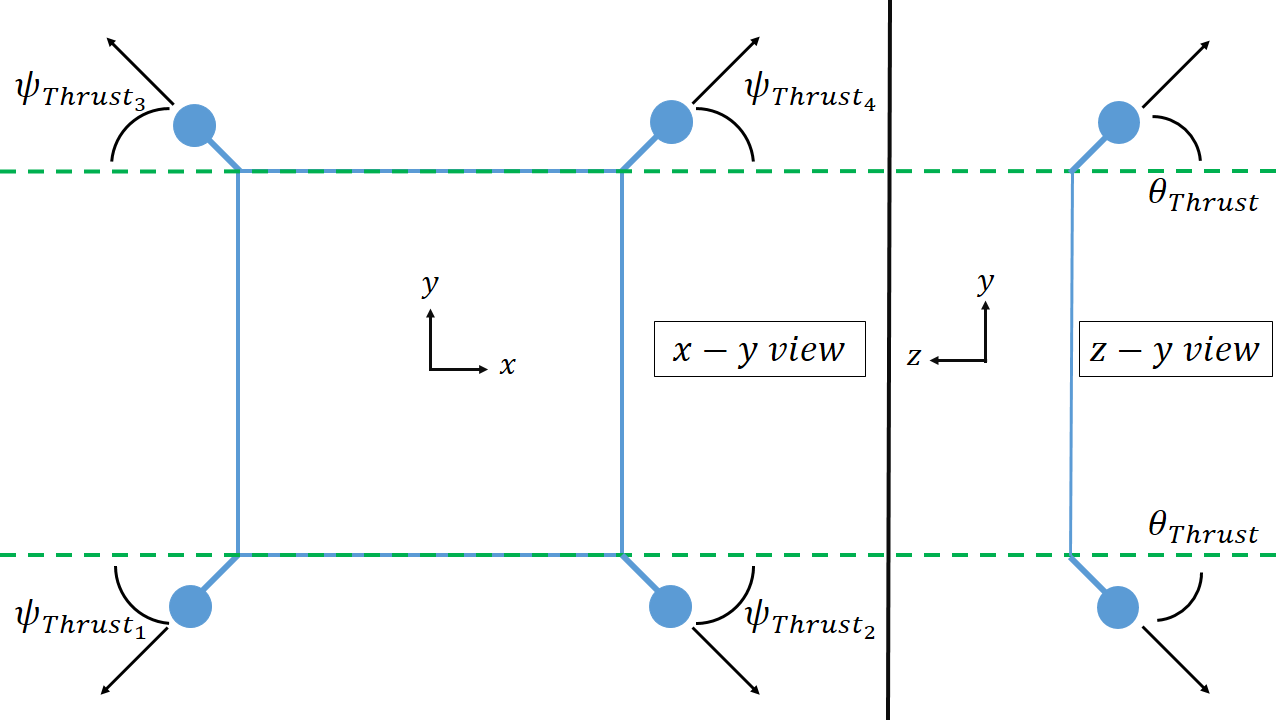}
    \caption{Thrust angles $\psi_{Thrust,i}$ and $\theta_{Thrust}$ for the MUs.} 
    \label{fig:Tangles}
\end{figure*}

\begin{table}[ht!]
\caption{\label{tab:table1} Simulation Parameters.}
\centering
\begin{tabular}{lcccc}
\hline
Parameter&Value\\\hline
\multicolumn{5}{c}{\textbf{Chaser, Tether, and Winch}}\\
\hline
Chaser Side Length $L_{ch}$, m & 1.5\\
Chaser Mass $m_{ch}$, kg & 1600\\
Chaser and Net Initial Distance $d_{ch}$, m & 0.1\\
Winch Radius $r_{w}$, m & 0.05\\
Winch Height $h_{w}$, m & 0.02\\
Winch Mass $m_{w}$, kg& 0.1\\
Tether Radius $r_{t}$, m & 0.002\\
Tether Density $\rho_{t}$, kg/m$^3$& 1390\\
Tether Young's Modulus $E_{t}$, GPA& 70\\
Tether Axial Damping Ratio $c_{t}$& 0.106\\
\hline
\label{tab:nominalparams}\\
\end{tabular}
\quad
\begin{tabular}{lcccc}
\hline
Parameter&Value\\\hline
\multicolumn{5}{c}{\textbf{Net, Corner Threads, and MU}}\\
\hline
Side Length $L_{net}$, m& 22.0\\
Mesh Length $l_{net,0}$, m& 1.0\\
Thread Radius $r_{net}$, m& 0.0011\\
Thread Density $\rho_{net}$, kg/m$^3$& 1390.0\\
Net Young's Modulus $E_{net}$, GPA& 70\\
Net Axial Damping Ratio $c_{net}$& 0.106\\
Corner Thread Length $l_{CT,0}$, m& 1.4142\\
Corner Thread Radius $r_{CT}$, m& 0.0007 \\
MU radius, $r_{MU}$, m& 0.0605 \\
\hline
\label{tab:nominalparams}\\
\end{tabular}
\begin{tabular}{lcccc}
\hline
Parameter&Value\\
\hline
\multicolumn{5}{c}{\textbf{Initial Conditions and Timestep}}\\
\hline
MU Ejection Speed $v_{e}$, m/s& 2.5\\
MU Shooting Angle $\theta_e$, deg &     36.87 \\
Net Stowed Side Length $L_{net,0}$, m  &     1.1 & \\
Simulation Time Step $\Delta$t, s & 10$^{-2}$\\
\hline
\end{tabular}
    
\label{tab:net}
\end{table}

\section{Formulation of the Optimization Baselines and RL Problems}\label{LearningtheOptimalApproachingandClosing Policies}

\subsection{Simulation Setup and the CQI}

The interactions with and modifications of the Vortex Studio-based simulator are done through a C++ Application Programming Interface
(API). The user defines net and target parameters, such as net thread radius, shooting angle, and the rotation speed of the target debris, in multiple .txt files as the inputs into the simulator. This work uses the Python programming language to implement the \ac{RL} component, while MATLAB implements the optimization component. 

To determine the effectiveness of the system in a scenario in which a great number of simulations is necessary for the optimization and \ac{RL} task, a quantitative metric referred to as the CQI is utilized \cite{CQI_Original, barnes}. The CQI value considers the similarity between the convex hull shape of the net and the target and net-target center of mass distance (COM) and is mathematically defined as:

\begin{equation}
    J_{n} = 0.1\frac{|V_n-V_t|}{V_t}+0.1\frac{|S_n-S_t|}{S_t} +0.8\frac{|q_n|}{L_c}
    \label{eq:cqiSafe}
\end{equation}

\noindent where the CQI at the $n$-th time-step, the convex hull (CH) volume of the net at the $n$-th time-step, the volume of the target, the CH surface area of the net at the $n$-th time-step, the surface area of the target, the distance from the center of mass of the target to the net’s COM at the $n$-th time-step, and the characteristic length of the target, defined as the shortest distance from the target’s COM to its surface and represented as $J_{n},$ $V_n,$ $V_t,$ $S_n,$ $S_t,$ $q_n,$ and $L_c$ respectively. Barnes and Botta's version of the CQI has been shown to effectively classify successful and unsuccessful captures \cite{barnes}. The target chosen for this paper is the second stage of the Zenit 2 launch vehicle (see Fig. \ref{fig:zenitShowcase}), which was also the subject of previous works utilizing Vortex Studio \cite{barnes,bayesopt}. The target has a mass of 9000 kg and dimensions of 3.9 m in diameter and 11.0 m long. As such, the values $V_t$, $S_t$, and $L_c$ associated with the target are 125.3 m$^3$, 159.9 m$^2$, and 1.95 m respectively.

\begin{figure}[h]
    \centering
    \includegraphics[width=0.49\textwidth]{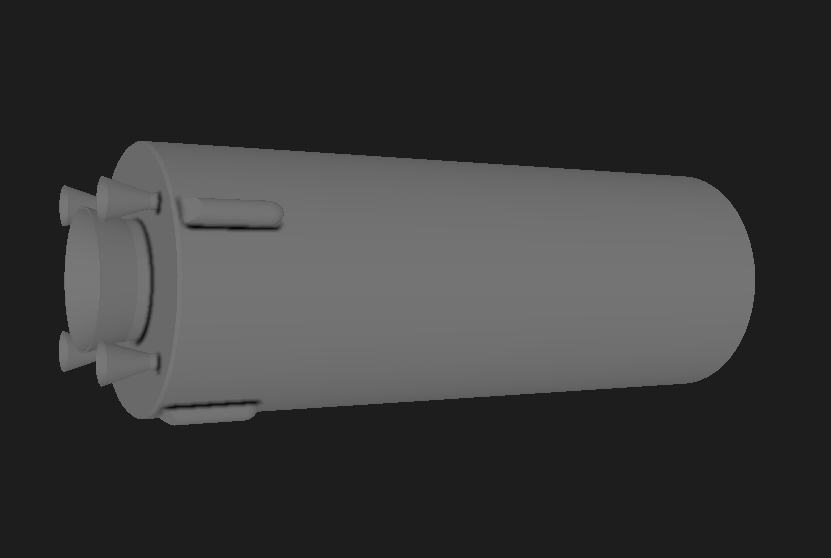}
    \caption{Model of the second stage of the Zenit 2 launch vehicle}
    \label{fig:zenitShowcase}
\end{figure}

The simulation scenario includes two phases: deployment and capture. In the deployment phase, the net leaves the chaser with an initial velocity and takes approximately 15.0 s to reach an almost fully-expanded state. When the near fully-expanded configuration is reached, the thrusters are activated. Thrusters remain activated until the net's center of mass reaches the closing distance from the target's center of mass, set to be 2.5 m. Once the distance is reached, the system enters the capture phase. The closing mechanism is triggered at the beginning of this phase, which applies constant forces on each pair of adjacent nodes of the closing threads, thus closing the net mouth. This phase is set to last 20.0 s after closing mechanism activation, and in the end, a settled CQI $I^*_{CQI}$, and the number of locked node pairs of the closing mechanism $N_L$ are returned. After the simulation, a .txt file containing the two output values and fuel consumed during the mission is generated. The trajectory of the net can be established and observed by creating an animation based on saved screenshots from the simulation.

\subsection{Defining the Optimization Task}
As previously mentioned, problems regarding realism need to be considered. For example, because sensing and observation cameras cannot be placed at the exact center on the face of the chaser where the net is launched due to the net ejection mechanisms, the axis of the launch and the target debris may not be perfectly aligned. Therefore, an offset of the target's position should be considered. In the RMSN model, the thrusters on the MUs can correct the trajectory of the net to address the offset mid-deployment. During this mission, the thruster angles must be considered. An optimization case study is designed with the angles as action variables. The objective is to successfully capture the target with minimal fuel consumption with varying action variables. To test the capability of the design to capture a target with a nonzero offset, the target's initial position is set to have a 9.0 m offset on the X-axis. The value of 9.0 m is much greater than expected in reality. However, it is chosen to demonstrate the system's adaptability with thrusters on each MU.

During this mission, three parameters must be considered: thruster angles, thrust magnitude, and the initial mass of the MUs. Three optimization case studies are designed with these three parameters as action variables. All three case studies aim to successfully capture the target with minimal fuel consumption with varying action variables. The target's initial position is set to have a 9.0 m offset on the X-axis to test the capability of the design to capture a target with a nonzero offset. 

The optimization method used in this paper is Bayesian Optimization \cite{bayesopt}, which has been successfully applied in various fields, including hyperparameter tuning for machine learning models, robotics, and experimental design. The acquisition function used for Bayesian Optimization in this research was Expected Improvement Plus. Compared to the vanilla version of the Expected Improvement \cite{EI} acquisition function, it can modify behaviors when an area is over-exploiting. 

 \emph{Case Study 1: Minimizing Fuel Consumption with Thrust Angles.} In this study, the objective is to find the minimum fuel consumption of the thrusters by only controlling the thruster angles $\psi_{Thrusts_i}$ and $\theta_{Thrusts}$. The initial mass is set to be 2.5 kg. The objective function is shown in Eq. \eqref{eq:obj_opt_1}. 

\begin{equation}
\label{eq:obj_opt_1}
\begin{aligned}
\min_{\mathbf{X}} \quad &  f_1(\mathbf{X})=m_p(\mathbf{X})\\
{s. t.}\quad & \mathbf{X}\in[\mathbf{X}_L,\mathbf{X}_U]\\
& g_1 = I^*_{CQI} \leq 2.5\\
& g_2 = N_L \geq 8\\
& g_3 = m_{f}\geq 2.0\\
{\text{where}:}\quad & \mathbf{X} = [\psi_{Thrust_1}, \psi_{Thrust_2}, \psi_{Thrust_3}, \\
& \psi_{Thrust_4}, \theta_{Thrust}]\\
\end{aligned}
\end{equation}
Where $m_p$ is the mass of the fuel consumed in each MU for each simulation, $\mathbf{X}$ represents the action variables picked from Table \ref{tab:designVar}, in which only the thrust angles $\psi_{Thrust_i}$ and $\theta_{Thrust}$ were chosen in this case study, $m_{f}$ represents the final mass of each MU when the thrusters shut down, which also equals to the mass of each MU at the end of the simulation. In this research, a successful capture threshold is set to be $I^*_{CQI}$ as 2.5, $N_L$ to be 8, and the mass of each MU at the end of the simulation, $m_{f}$ needs to be greater than 2.0 kg.
     
 \emph{Case Study 2: Minimizing Fuel Consumption with Thrust Angles and Initial MU Mass.} This study follows the previous case study's framework, but the initial mass of each MU is also used as one of the action variables. This case study aims to explore the optimal design of the initial mass and the thrust angles to minimize fuel consumption. The objective function is shown in Eq. \eqref{eq:obj_opt_2}:
 
\begin{equation}
\label{eq:obj_opt_2}
\begin{aligned}
\min_{\mathbf{X}} \quad &  f_1(\mathbf{X})=m_p(\mathbf{X})\\
{s. t.}\quad & \mathbf{X}\in[\mathbf{X}_L,\mathbf{X}_U]\\
& g_1 = I^*_{CQI} \leq 2.5\\
& g_2 = N_L \geq 8\\
& g_3 = m_{f}\geq 2.0\\
{\text{where}:}\quad & \mathbf{X} = [\psi_{Thrust_1}, \psi_{Thrust_2}, \psi_{Thrust_3},\\
&\psi_{Thrust_4}, \theta_{Thrust}, m_0]\\
\end{aligned}
\end{equation}
\noindent where $m_0$ is the initial mass of each MU.

\emph{Case Study 3: Minimizing Fuel Consumption with Thrust Angles, Magnitude, and Initial MU Mass.} This case study chooses the magnitude of the thrust force as an additional action variable. The fuel consumption rate is set to be proportional to the magnitude of the thrust force. This case study explores the optimal design of the initial mass, thrust magnitude, and thrust angles to minimize fuel consumption. The objective function is shown in Eq. \eqref{eq:obj_opt_3}:
\begin{equation}
\label{eq:obj_opt_3}
\begin{aligned}
\min_{\mathbf{X}} \quad &  f_1(\mathbf{X})=m_p(\mathbf{X})\\
{s. t.}\quad & \mathbf{X}\in[\mathbf{X}_L,\mathbf{X}_U]\\
& g_1 = I^*_{CQI} \leq 2.5\\
& g_2 = N_L \geq 8\\
& g_3 = m_{f}\geq 2.0\\
{\text{where}:}\quad & \mathbf{X} = [\psi_{Thrust_1}, \psi_{Thrust_2}, \psi_{Thrust_3}, \\
&\psi_{Thrust_4}, \theta_{Thrust}, m_0, F_{Thrust}]\\
\end{aligned}
\end{equation}
\noindent where $F_{Thrust}$ is the magnitude of the thrust force.

Table \ref{tab:designVar} summarizes the action variables and their bounds for the optimization tasks. The values for $\psi_{Thrust_i}$ for $i$ = 1, 2, 3, 4 and $\theta_{Thrust}$ are assigned a bound after initial manual tuning. This allows the optimization algorithm to search for optimal values close to what is already known to yield a feasible solution. The range of possible $m_0$ values for each MU is chosen to be approximately the same as the mass of a 2U CubeSat \cite{nieto2019cubesat}, which has a similar size to what each MU is envisioned to possess. Meanwhile, the range of $F_{Thrust}$ is chosen to be $\pm$3 N from the nominal thrust value.

\begin{table}[h]
	\begin{center}
		\caption{Bounds of Optimization Action Variables}
		\label{tab:designVar}
		\begin{tabular}{llll}
			\toprule
			Action Variables & Data Type  & Bounds & Step Size\\
			\midrule
            $\psi_{Thrust_1}$ &  Scalar & 70 to 90 deg & 0.1 deg\\
            $\psi_{Thrust_2}$ &  Scalar & 35 to 55 deg & 0.1 deg\\
            $\psi_{Thrust_3}$ &  Scalar & 70 to 90 deg & 0.1 deg\\
            $\psi_{Thrust_4}$ &  Scalar & 35 to 55 deg & 0.1 deg\\
            $\theta_{Thrust}$ &  Scalar & 35 to 55 deg & 0.1 deg\\
            $m_{0}$ & Scalar & 2.0 to 2.5 kg & 0.001 kg\\
            $F_{Thrust}$ & Scalar & 5 to 12 N & 0.0001 N \\
            
 			\bottomrule 
		\end{tabular}
	\end{center}
\end{table}
\subsection{Defining the Learning Task}
\ac{RL} models the actions as Markov Decision Processes (MDP) \cite{mdp}. The objective is to capture the target debris and minimize fuel consumption successfully. The actions in this model are the thrust angles, which activate at the time step of 15.0 s. The simulation shuts down the thrusters when the closing condition is met. To test the generalization of the design, a uniformly distributed noise with the range of (-5.0, 5.0) m is added to the target's Z-direction initial position. Therefore, the MDP of this problem can be simplified, where the target parameters define the state space, and the thrusters' angles define the action space. The details are shown in Table \ref{tb:net2} and Table \ref{tb:action_space}. The state space in the current framework has five parameters, but only Z-axis Offset is the changing parameter. The rest four parameters are fixed and kept in the framework for future study of \ac{RL} by adding more complexity to the target's position, orientation, and angular velocity magnitude.
\begin{table}
	\begin{center}
		\caption{Parameters of State Space}
		\label{tb:net2}
		\begin{tabular}{lll}
			\toprule
			Target Parameters  & Data Type & Bounds\\
			\midrule
            Z-axis Offset & Scalar & -45 to -55 m \\
			\bottomrule 
		\end{tabular}
	\end{center}
	
\end{table}

\begin{table}
	\begin{center}
		\caption{Parameters of Action Space}
		\label{tb:action_space}
		\begin{tabular}{lll}
			\toprule
			Thrusters Parameters  & Data Type  & Bounds\\
			\midrule
            $\psi_{Thrust,1}$ &  Scalar & 70 to 90 deg\\
            $\psi_{Thrust,2}$ &  Scalar & 35 to 55 deg\\
            $\psi_{Thrust,3}$ &  Scalar & 70 to 90 deg\\
            $\psi_{Thrust,4}$ &  Scalar & 35 to 55 deg\\
            $\theta_{Thrust}$ &  Scalar & 35 to 55 deg\\
			\bottomrule 
		\end{tabular}
	\end{center}
\end{table}
The actions and states are sent to the simulator, and the results of the simulation are used for the calculation of the reward function:
\small
\begin{equation}
    \label{eq:reward}
        \begin{aligned}
        \max_{\mathbf{Q}}\quad & R = r_{\text{fuel}} + r_{\text{CQI}}+ r_{\text{NL}} + r_{\text{mass}} + r_{\text{end}}\\
         \text{where: }\quad & r_{\text{fuel}} = m_0 - \lambda \cdot 1.2\cdot (t_{\text{sim}}-15-20)\\
         & r_{\text{CQI}} =
            \begin{cases}
             -\ln((I^*_{CQI}-2.5)^2+1), & \text{if } I^*_{CQI} > 2.5 \\
             0,              & \text{otherwise}
            \end{cases}\\
        & r_{\text{NL}} =
            \begin{cases}
            -\ln((N_L-8)^2+1), & \text{if } N_L < 8 \\
             0,              & \text{otherwise}
            \end{cases}\\
        & r_{\text{mass}} =
            \begin{cases}
             -\ln((m_{f}-2)^2+1), & \text{if } m_{f}<2 \\
             0,              & \text{otherwise}
            \end{cases}\\
        & r_{\text{end}} =
            \begin{cases}
             10, & \text{if} I^*_{CQI}\leq2.5 \wedge N_L\geq8 \wedge m_{f}\geq2\\
             0,              & \text{otherwise}
            \end{cases}\\            
        \end{aligned} 
\end{equation}
\normalsize

\noindent where $\mathbf{Q}$ represents the policy model; $R$ represents the reward in every episode; $r_{\text{fuel}}$ represents the reward based on fuel consumption on each thruster; $m_0$ and $m_{f}$ are the initial and end mass of each MU; $\lambda$ is the fuel burning rate; the constant 1.2 is to add twenty-percent more consumption of the fuel as a safety factor; the constant 15 and 20 are the fixed time cost to wait for the net to expand fully and for the CQI to settle; $t_{\text{sim}}$ is the total simulation time; $r_{I^*_{CQI}}$, $r_{\text{NL}}$, $r_{\text{mass}}$ are the rewards based on the CQI, number of locked pairs, and MUs at the end of the simulation, which act as the logarithmic penalty function to penalize the reward if the constraints are violated; $r_{\text{end}}$ is the terminal reward.

In this paper, the objective of \ac{RL} is to find the optimal trajectory of the corner MUs and the energy cost to capture the target. The thrust angles determine the trajectory and the energy cost, defined as fuel consumption during the approaching phase. The reward received is defined to be the mass of the remaining fuel after the approaching phase ends. The conditional formulations in the reward function, $r_{\text{CQI}}$, $r_{\text{NL}}$ and $r_{\text{mass}}$, ensure essential penalties for missing the target (too large settled CQI), insecure capture (too few locked pairs) and consuming too much fuel (remaining mass is less than the dry mass). The logarithmic penalty functions also ensure the penalty is not too large, which could jeopardize the learning because the settled CQI can reach a value of several hundreds for a failed capture. The terminal bonus state reward, $r_{\text{end}}$, is for the capture that does not reach any of the penalty states, which can prevent the policy model from exploiting only one of the penalty states, such as minimizing the settled CQI or only maximizing the number of locked pairs.

The learning technique used in this framework is Proximal Policy Optimization (PPO) \cite{ppo} from stable baselines3 \cite{stable-baselines3}. It is a \ac{RL} algorithm that combines the benefits of trust region policy optimization (TRPO) \cite{trpo} and traditional policy gradient methods. It is designed to balance exploration and exploitation while training deep neural networks for optimal policy learning. PPO builds upon gradient methods by introducing a surrogate objective function with a clipped probability ratio, ensuring the policy updates are limited to a trust region around the old policy. This prevents excessively large policy updates that can destabilize the training process. It also divides the data received into smaller batches and updates the policy parameters incrementally. This approach provides a more efficient and computationally tractable method for optimizing the policy, as it reduces the variance of gradient estimates and allows for more frequent updates. In PPO, the policy network outputs a probability distribution over actions, which is used to sample actions during training and evaluation. The value network estimates the expected cumulative reward from a given state, which is used for temporal difference learning to update the value function and policy. The neural network used in this research is a multi-layer perceptron (MLP) \cite{mlp} with 2 layers and 64 neurons.


\section{Simulation Results}
\subsection{Optimization Case Study 1 and Reinforcement Learning Results}

For optimization Case Study 1, the hardware used is a Windows workstation with an AMD Ryzen 9 5950X 16-Core Processor and 64 GB RAM. The time cost was 17 hours, with five action variables, 500 iterations, 200 points in the active set, and 80 initial sampling points. 
The minimum fuel consumption value and the action variables at the minimum fuel consumption are shown in the first row of Table \ref{tab:case1_RL}. Figure \ref{fig:opt50} showcases selected instances from the optimal Case Study 1 simulation. Figure \ref{fig:opt50}(a) shows the configuration of the system in the instance the thrusters are activated. Meanwhile, Fig. \ref{fig:opt50}(b) and (c) highlight the motion of the tether-net under propulsion as it proceeds towards the target. Lastly, Fig. \ref{fig:opt50}(d) shows the net as it wraps around the target.

\begin{figure}[ht!]
\centering
\begin{subfigure}{.49\linewidth}
  \includegraphics[width=1\linewidth]{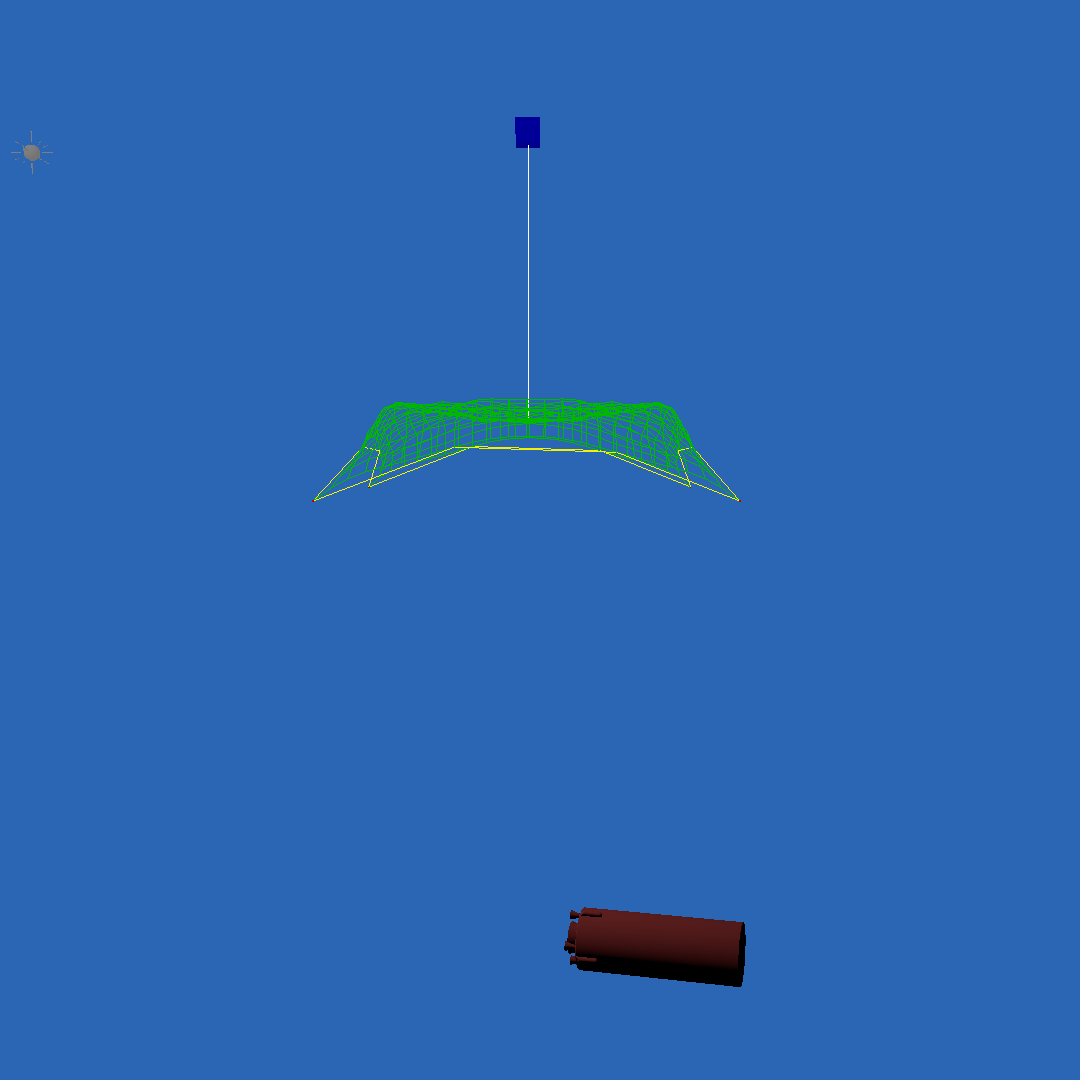}
 \subcaption{$t = 15.0$ s}
\end{subfigure}%
\begin{subfigure}{.49\linewidth}
  \includegraphics[width=1\linewidth]{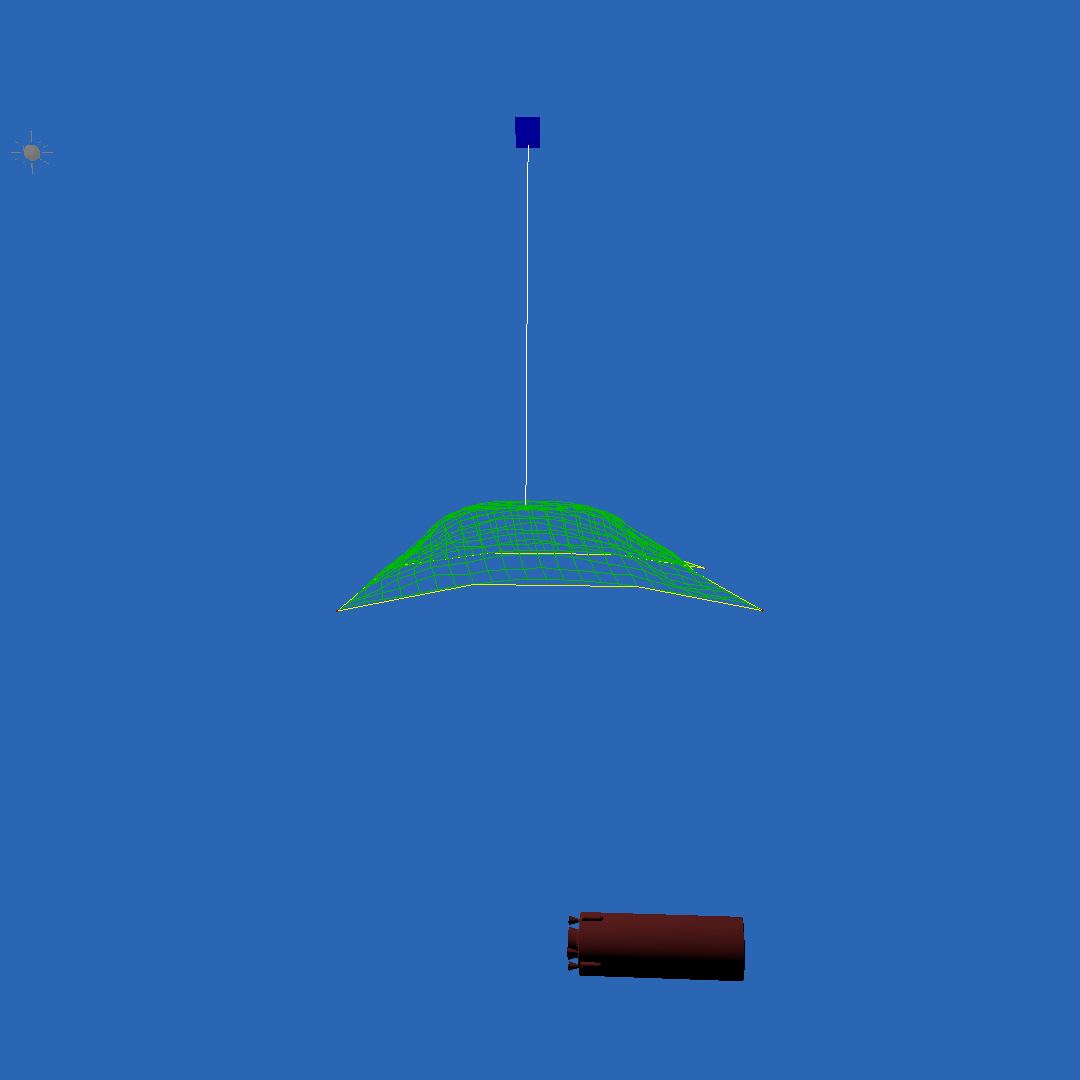}
   \subcaption{$t = 17.0$ s}
\end{subfigure}
\begin{subfigure}{.49\linewidth}
  \includegraphics[width=1\linewidth]{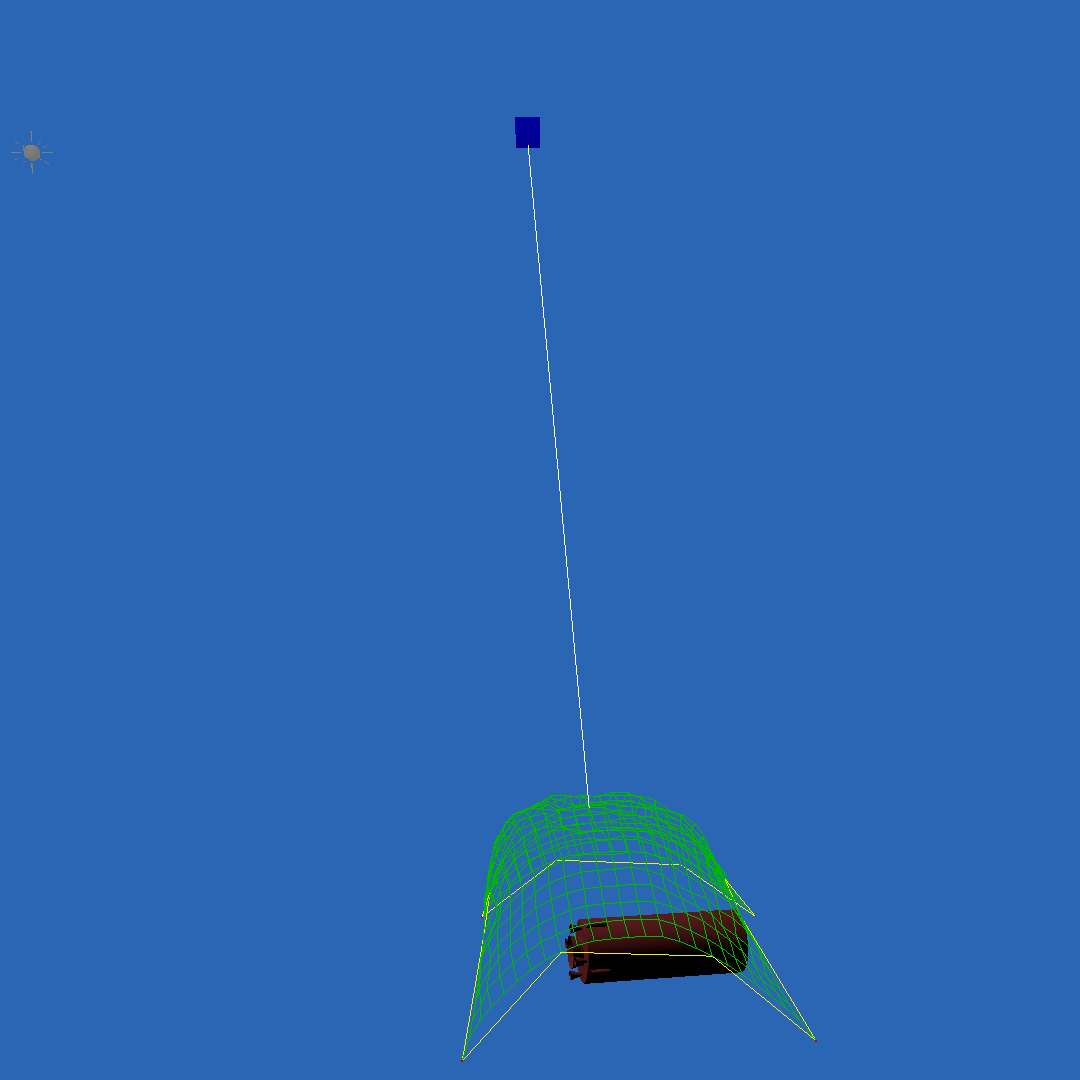}
 \subcaption{ $t = 20.0$ s}
\end{subfigure}%
\begin{subfigure}{.49\linewidth}
  \includegraphics[width=1\linewidth]{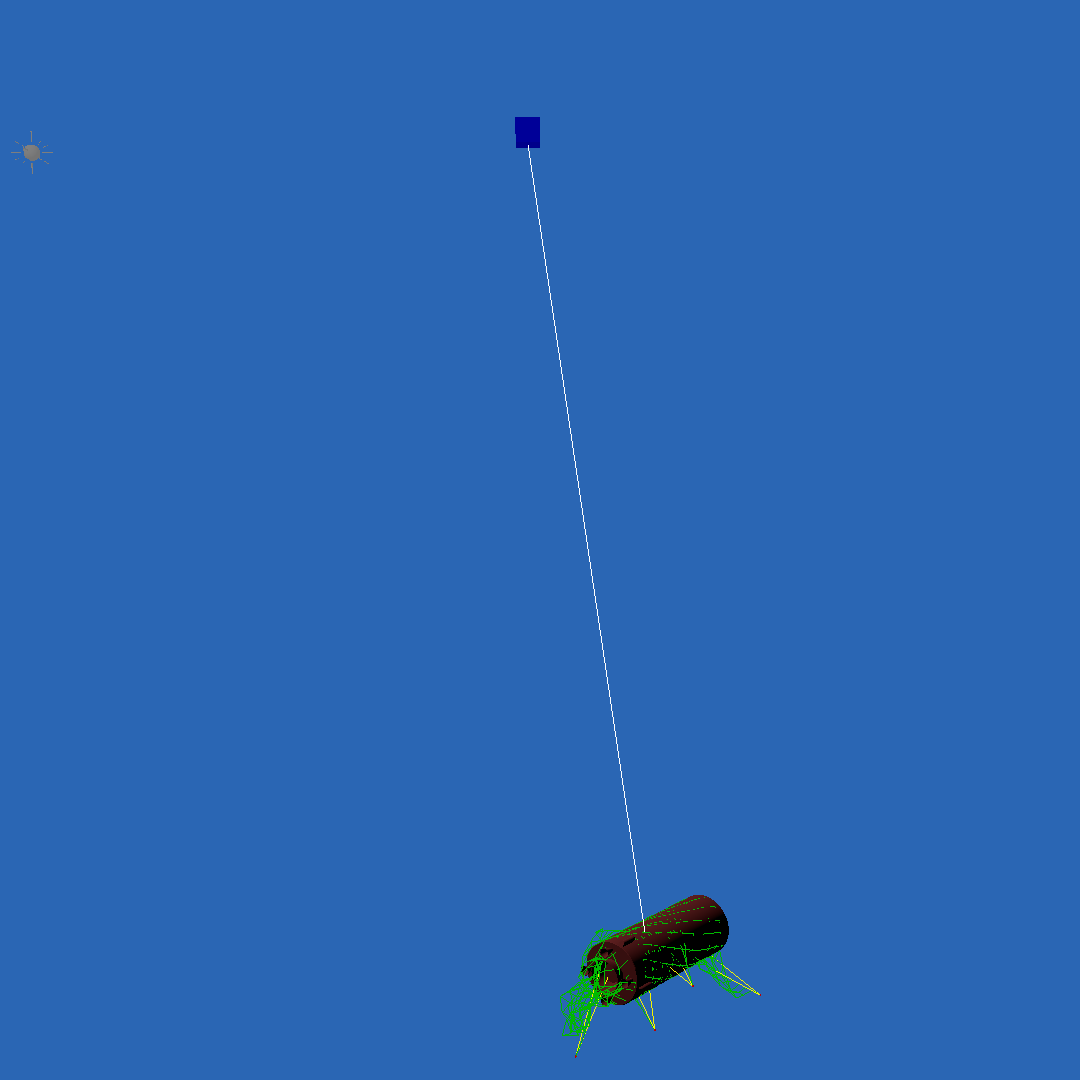}
   \subcaption{ $t = 28.0$ s}
\end{subfigure}
\caption{Optimization Capture Simulation at -50 m Z-axis position}
\label{fig:opt50}
\end{figure}

Meanwhile, \ac{RL} was performed on a Windows workstation with an AMD Ryzen 9 5950X 16-Core Processor and 64 GB RAM. The environment was vectorized for parallel training of 32 episodes simultaneously. The mini-batch size was 64. The learning rate was tuned during the process of training. For the first 1600 episodes, the learning rate was 0.0001. However, the plot of average reward showed no sign of learning and still had significant fluctuations. Therefore, the learning rate was increased to 0.001 for the next 8000 episodes, and the plot shows the average reward starts to increase. For the last 2112 episodes, the learning rate was reduced to 0.0005. Each episode has only one step in this \ac{RL} framework. The learning process has 11712 episodes and took 511 hours to finish. As the learning progresses, the episode takes longer because the successful capture requires more time to simulate in Vortex Studio.

Figure \ref{fig:reward} shows that the model was learning over time but still shows strong fluctuations and has not converged at the end. For a successful capture, the reward is over 12, and if the capture misses the target, the reward is below -12. Figure \ref{fig:reward} shows the average reward for every 32 and 192 episodes, and the average reward was initially around 1, and in the end, it reached above 10. The trials with rewards in between usually violate one of the constraints, either the settled CQI or the number of locked pairs. The trend of the rewards with the fluctuations is a sign of insufficient training. The saved policy model was then tested with the same initial Z-axis position (-50 m) as the optimization result. The predicted thrust angle is shown in the second row of Table \ref{tab:fuel}, with total fuel consumption of 0.083 kg, and it did not violate the constraints of settled CQI and the number of locked pairs, which made a secure capture. Figure \ref{fig:RL50} shows instances within the capture simulation -- similar to Fig. \ref{fig:opt50} -- with the thrust angles defined by the policy model. This demonstrates that the policy model can capture a target with the same fixed position as the optimization one, and the fuel consumption of RL is only 0.004 higher.

\begin{table*}[h]
\begin{center}
	\begin{center}
 \caption{\label{tab:case1_RL}  Case Study 1 and RL Optimal Action Variables}
		\begin{tabular}{|l|l|l|l|l|l|}
			\hline
			Methods & $\psi_{Thrust_1}$, deg & $\psi_{Thrust_2}$, deg & $\psi_{Thrust_3}$, deg & $\psi_{Thrust_4}$, deg & $\theta_{Thrust}$, deg\\
			\hline
            Optimization Case Study 1 & 84.7  &  40.8  &  86  &  43.2  &  37.3\\
            \hline
            RL & 87.0  &  35.4  &  82.6  &  49.4  &  48.8\\
            \hline
		\end{tabular}
	\end{center}
\end{center}
\end{table*}

\begin{figure}[htp!]
    \centering
    \includegraphics[width=0.49\textwidth]{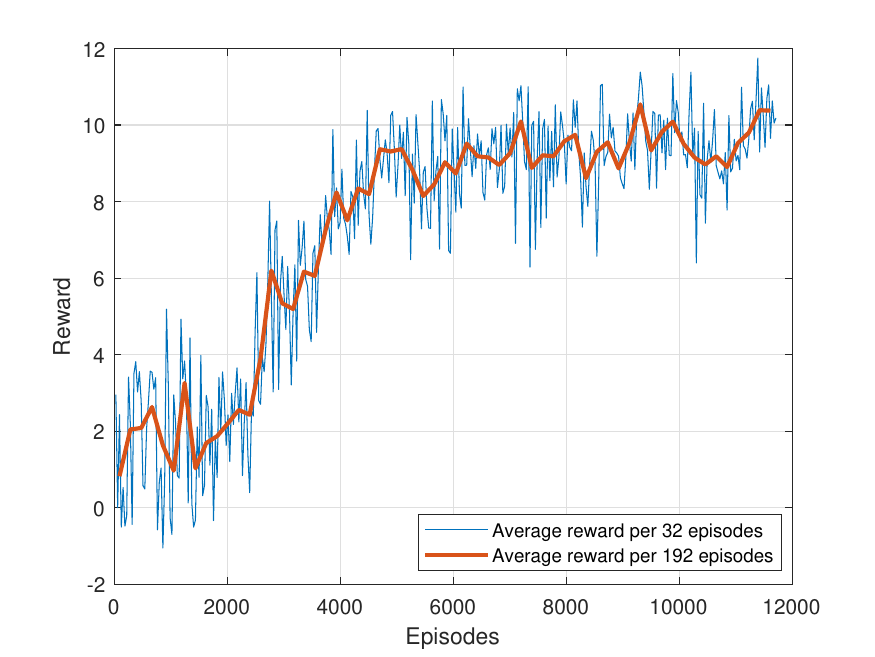}
    \caption{Average reward for the training of \ac{RL} policy model} 
    \label{fig:reward}
\end{figure}

\begin{figure}[h!]
\centering
\begin{subfigure}{.49\linewidth}
  \includegraphics[width=1\linewidth]{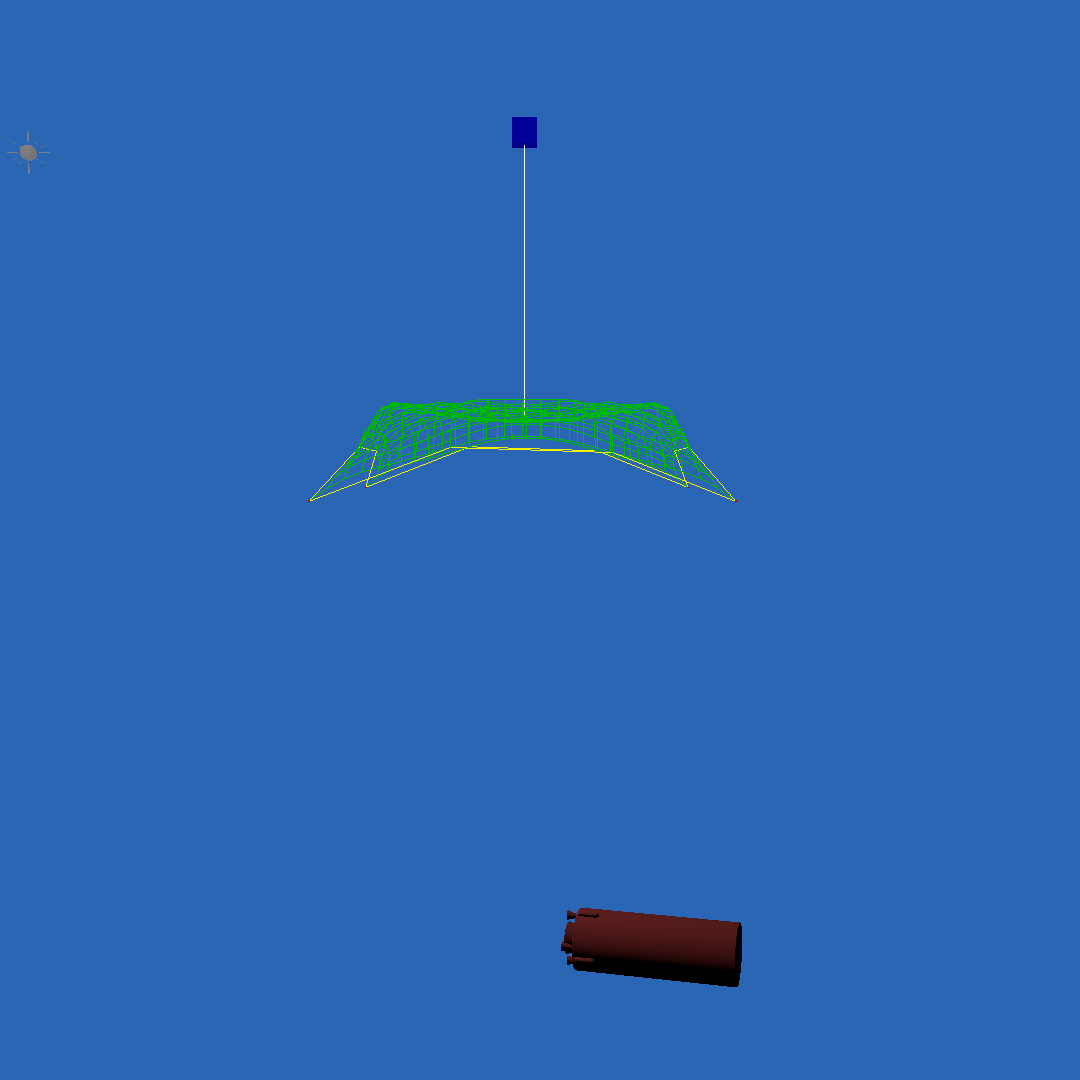}
 \subcaption{$t = 15.0$ s}
\end{subfigure}%
\begin{subfigure}{.49\linewidth}
  \includegraphics[width=1\linewidth]{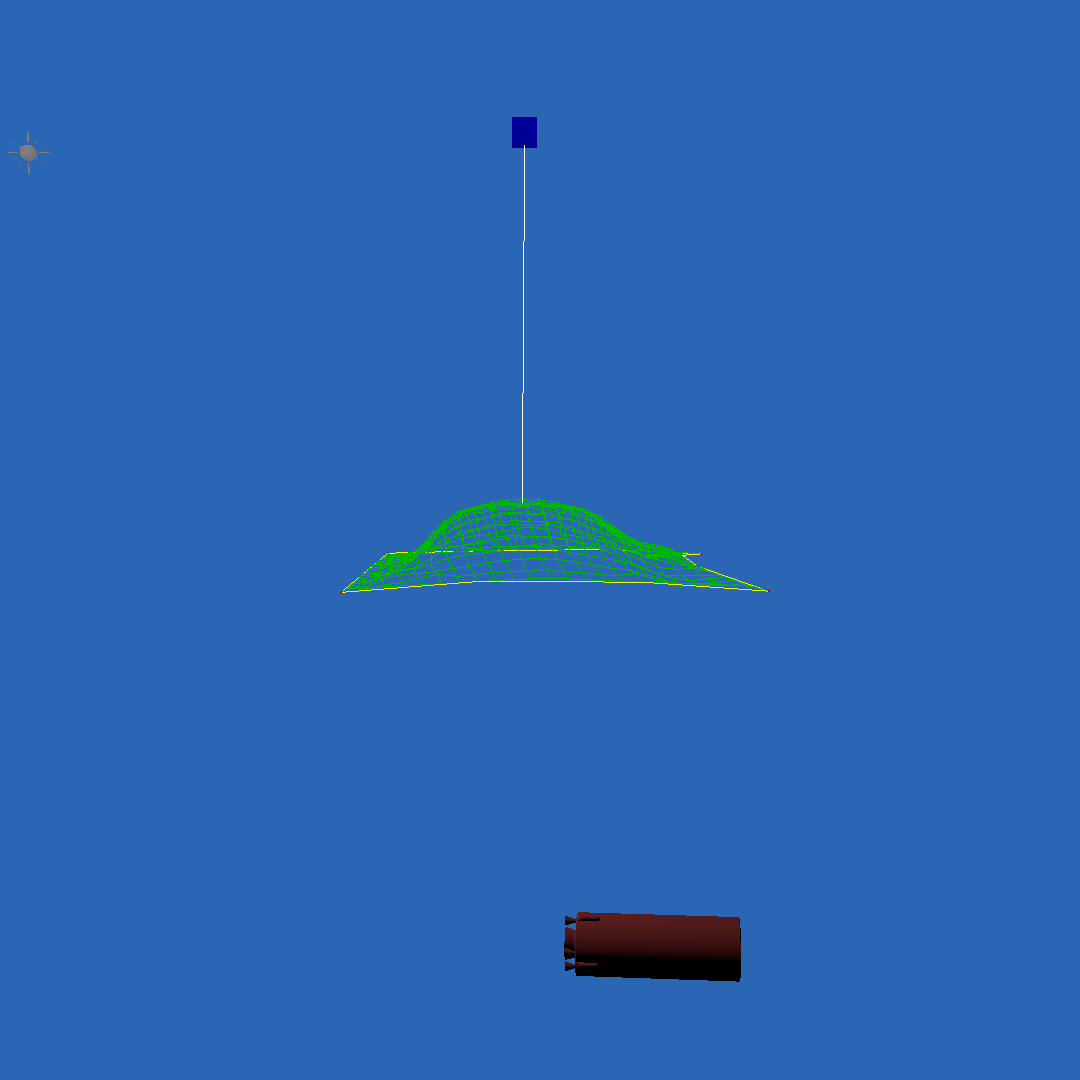}
   \subcaption{$t = 17.0$ s}
\end{subfigure}
\begin{subfigure}{.49\linewidth}
  \includegraphics[width=1\linewidth]{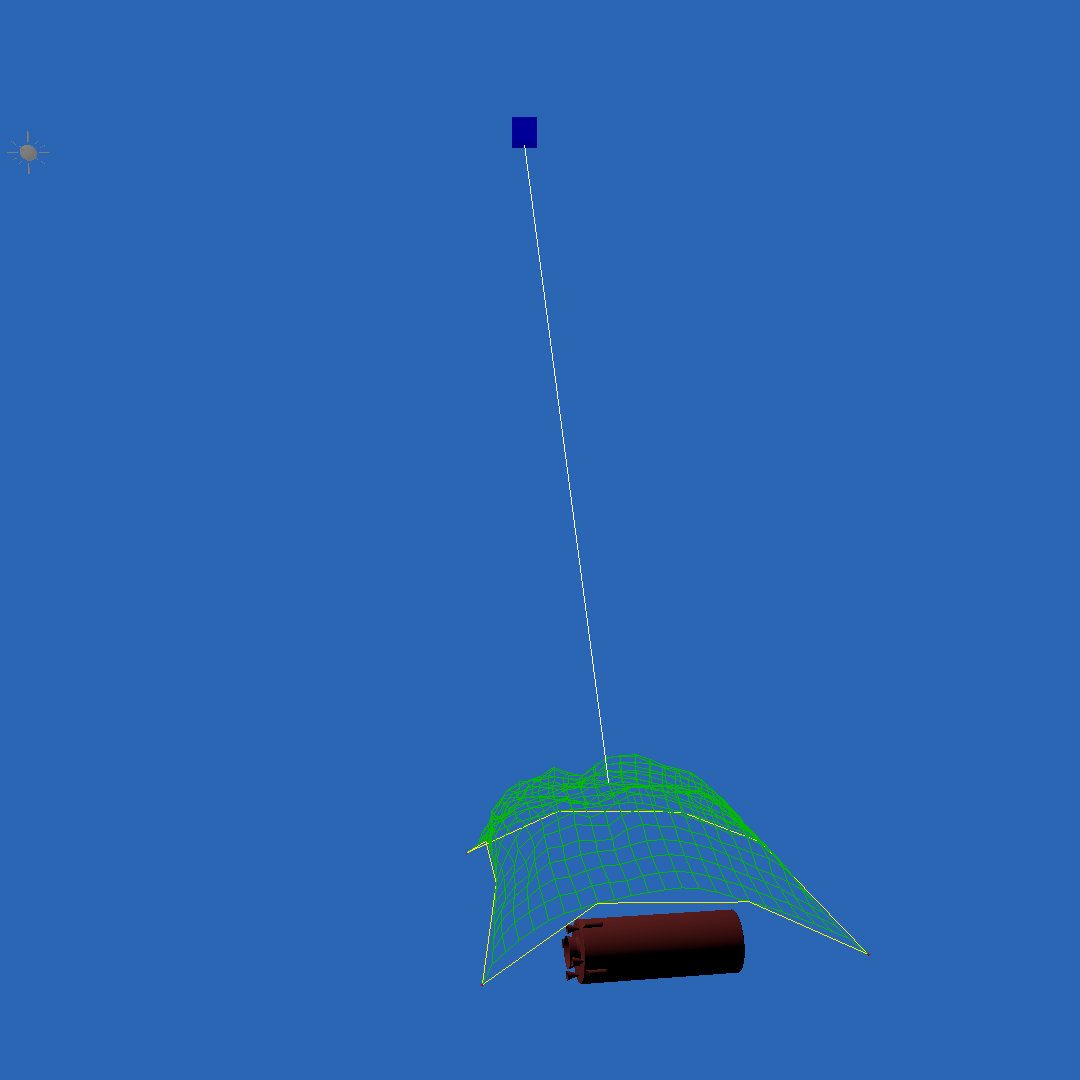}
 \subcaption{ $t = 20.0$ s}
\end{subfigure}%
\begin{subfigure}{.49\linewidth}
  \includegraphics[width=1\linewidth]{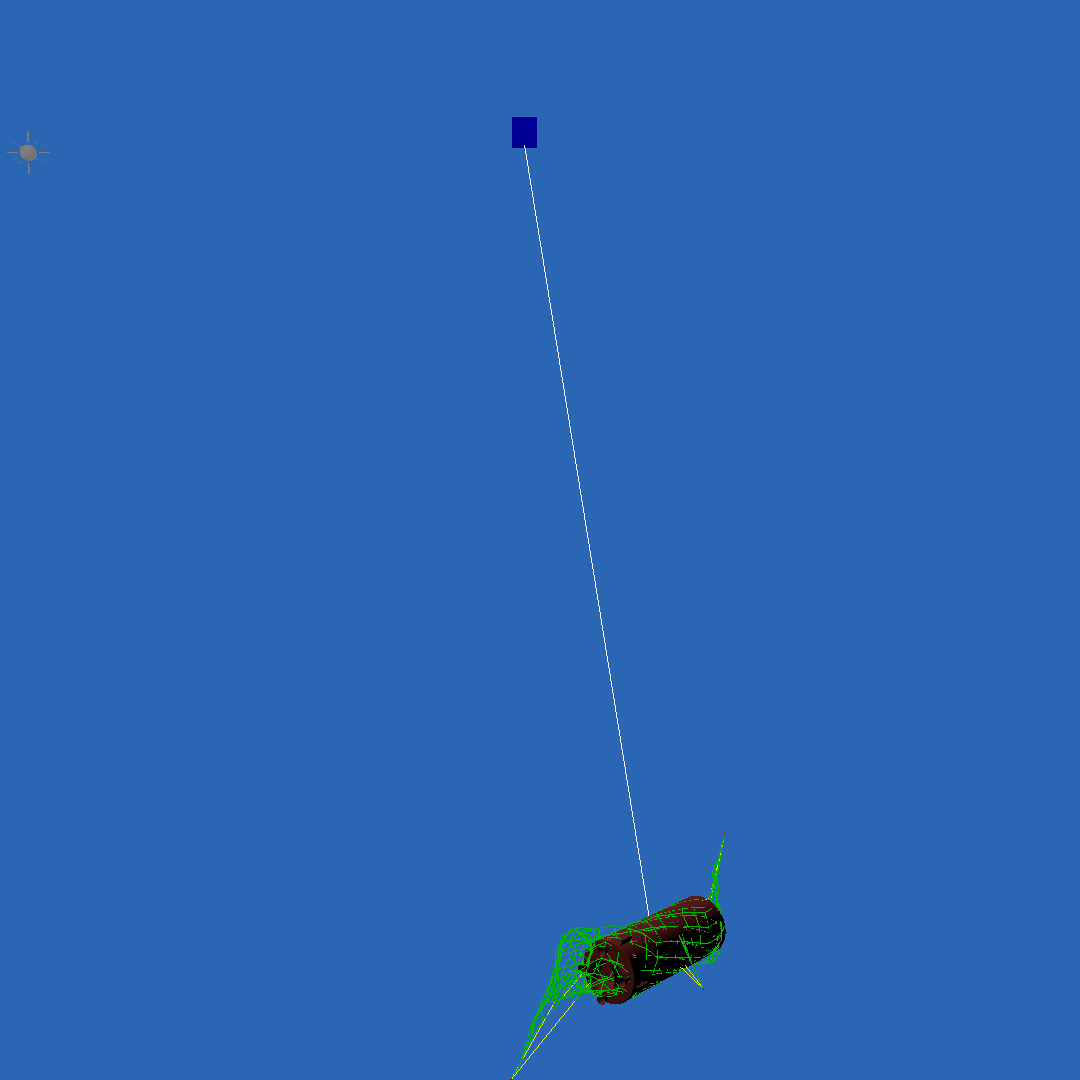}
   \subcaption{ $t = 28.0$ s}
\end{subfigure}
\caption{RL Capture Simulation at -50 m Z-axis position}
\label{fig:RL50}
\end{figure}

The performance of the \ac{RL} policy model and the optimization Case Study 1 -- both of which only affect the thrust angles of the MUs -- is tested with the noise in the range of -5.0 m to +5.0 m added to the target's initial Z-axis position. With the policy model, the predicted thrust angles change when the target's initial Z-axis position differs. At the same time, for the optimization Case Study 1 evaluation, they are fixed to be the values listed in the first row of Table \ref{tab:case1_RL}. The optimization results for this scenario were tested with 50 samples with noise added to the initial Z-axis position, and the success rate was 46\%. The \ac{RL} model with the highest average reward occurs in episode 11360. The model was tested with the same noise added as the optimization one, and the success rate of the 50 samples was 88\%. The distribution of the successful capture of the two cases is shown in Fig. \ref{fig:dist}(a) and \ref{fig:dist}(b). The plots show that the \ac{RL} model has a higher successful capture rate, and the successful captures are more evenly distributed across -45.0 m to -55.0 m for the Z-axis position of the target. Figure \ref{fig:dist}(c) shows that the optimization result's median successful capture Z-axis position was -51.35 m, and the reinforcement result's median successful capture position was -50.28 m.

\begin{figure}[ht!]
     \centering
     \begin{subfigure}[b]{0.4\textwidth}
         \centering
         \includegraphics[width=\textwidth]{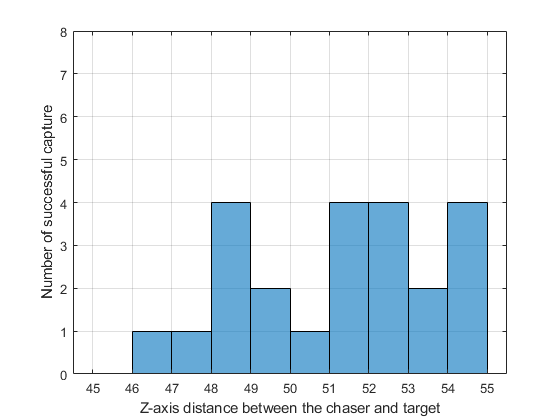}
         \caption{Number of successful capture with optimization result}
         \label{fig:dstr_opt}
     \end{subfigure}
     \hfill
     \begin{subfigure}[b]{0.4\textwidth}
         \centering
         \includegraphics[width=\textwidth]{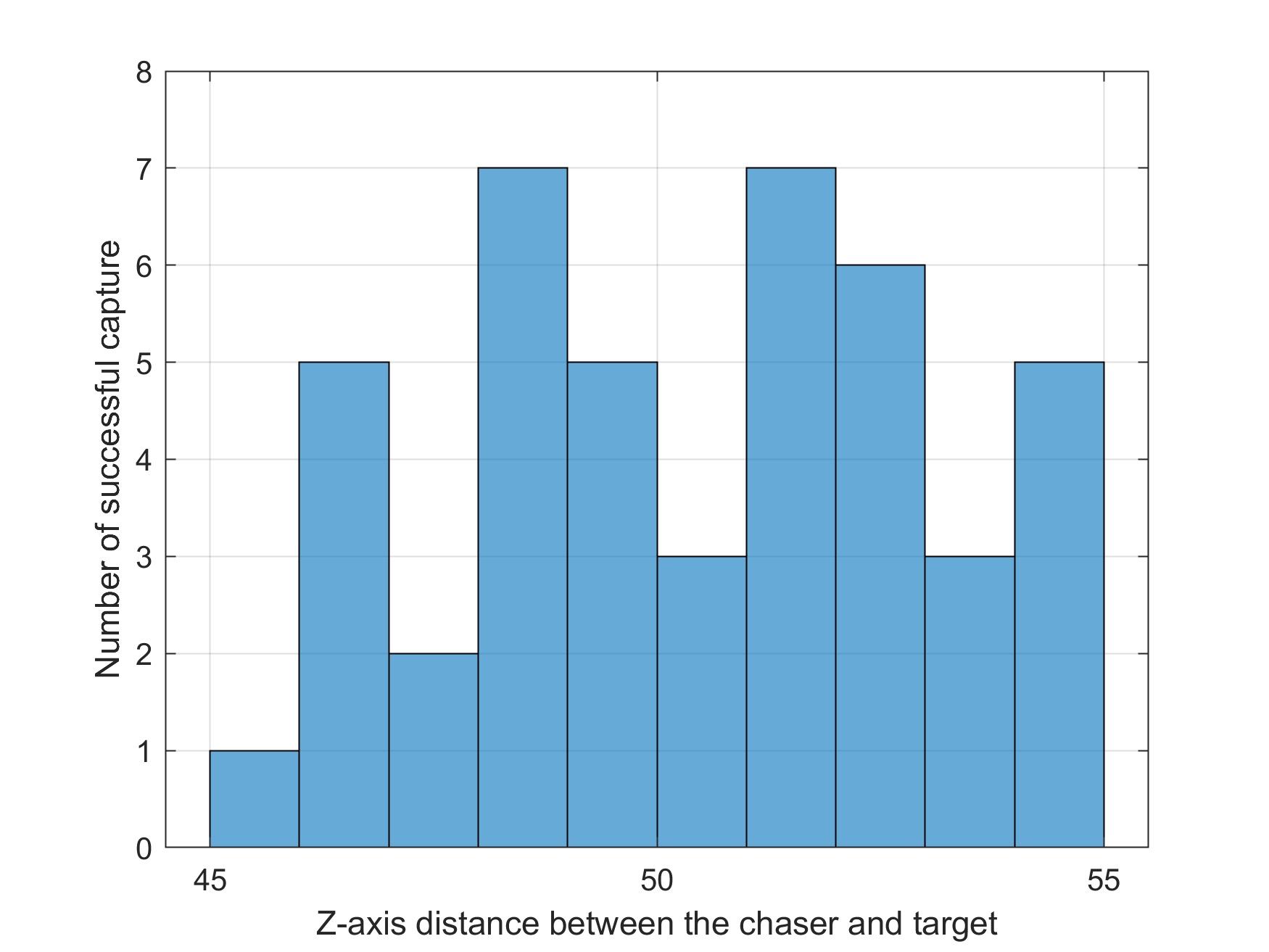}
         \caption{Number of successful capture with RL result}
         \label{fig:dstr_rl}
     \end{subfigure}
     \hfill
     \begin{subfigure}[b]{0.4\textwidth}
         \centering
         \includegraphics[width=\textwidth]{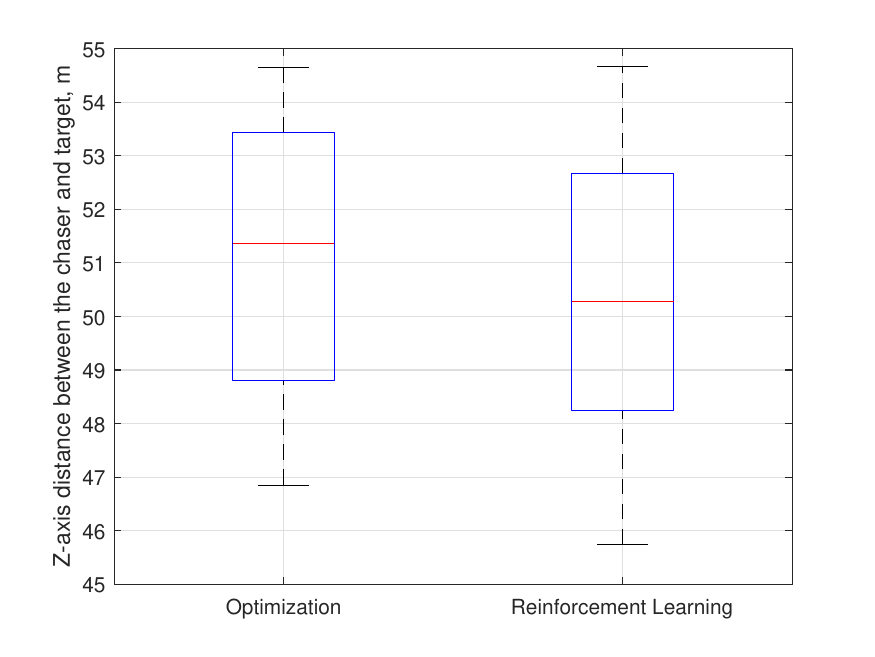}
         \caption{Boxplot of successful capture}
         \label{fig:boxplot}
     \end{subfigure}
        \caption{Case Study 1 Reinforcement Learning policy model and optimization result comparison}
        \label{fig:dist}
\end{figure}

Though the total learning time of the \ac{RL} model is over five hundred hours, once the policy model of \ac{RL} is trained, the execution time to predict the ideal thrust angles is only 41 milliseconds. The optimization method for one scenario takes 17 hours. Still, to improve the capture success rate, optimization needs to run in all 50 scenarios, which will take over eight hundred hours by estimation. Therefore, the \ac{RL} method has a more efficient generalized performance.

\subsection{Optimization Case Study 2 and 3}

For Case Study 2, using the same computer as Case Study 1, the optimization time cost was 24.1 hours, with 6 action variables, 500 iterations, 200 points in the active set, and 80 initial sampling points. This case study is intended to be compared with Case Study 1 optimization. The minimum fuel consumption value and the action variables at the minimum fuel consumption are shown in Fig. \ref{fig:opt_results}(b) and the first row of Table \ref{tab:case2_3}. The minimum objective value is smaller than that of Case Study 1, which shows that the fuel consumption can be lower with a smaller thrust magnitude for each thruster while the capture is still successful. Meanwhile, in Case Study 3, using the same computer as the previous two optimization cases, the time cost was 38.9 hours, with 7 action variables, 500 iterations, 200 points in the active set, and 80 initial sampling points. This case study aims to find the optimal actions of each MU's thrust angle, magnitude, and initial mass. For all three optimization case studies, the minimum fuel consumption values over function evaluations are shown in Fig. \ref{fig:opt_results}, and Table \ref{tab:fuel} displays the optimal fuel consumption for the 3 cases as well as the fuel consumption of the \ac{RL} policy model with the target possessing -50 m Z-axis position. Case Study 3 obtained the lowest fuel consumption of all 3 cases, demonstrating that tuning the initial mass of each MU -- in addition to the thrust angles and magnitude -- leads to additional fuel savings.

Table \ref{tab:fuel} compares minimum fuel cost at the -50 m scenario of the three optimization cases and RL and the training time and execution time of optimization and RL. Though the RL method has higher fuel consumption and takes longer to train, once the RL model is trained, the execution time is much shorter than the optimization cases. For the optimization method to get the result, it has to run the optimizing process; however, RL only needs to input the state to the policy model, and the result can be generated in less than a second.

\begin{figure}[ht!]
     \centering
     \begin{subfigure}[b]{0.4\textwidth}
         \centering
         \includegraphics[width=\textwidth]{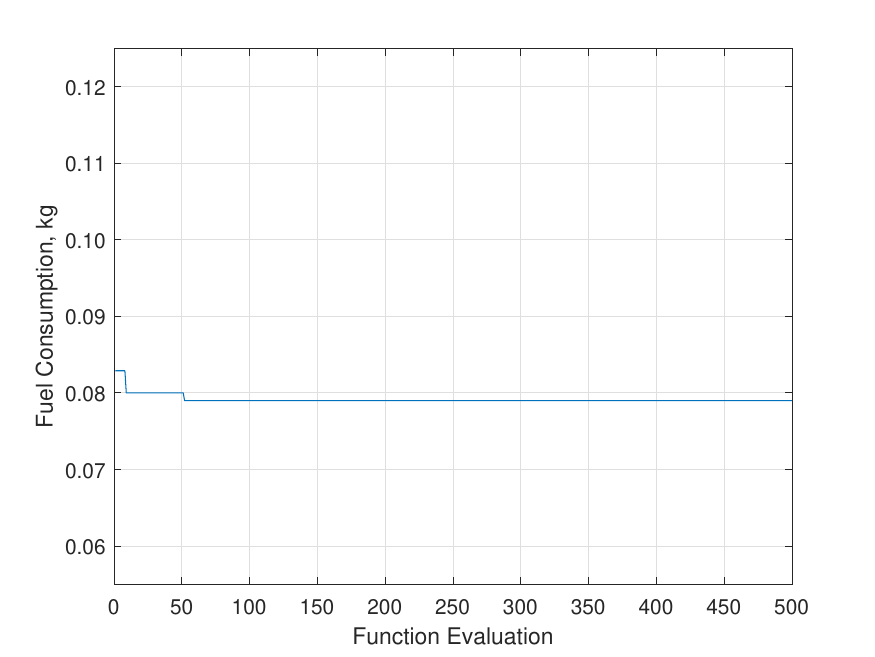}
         \caption{Case 1}
         \label{fig:opt_case1}
     \end{subfigure}
     \hfill
     \begin{subfigure}[b]{0.4\textwidth}
         \centering
         \includegraphics[width=\textwidth]{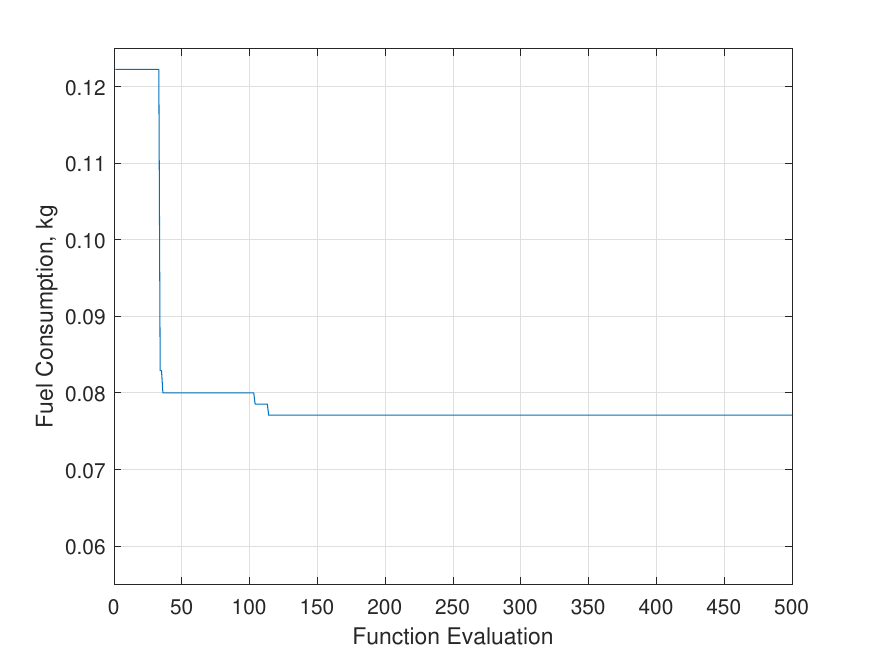}
         \caption{Case 2}
         \label{fig:opt_case2}
     \end{subfigure}
     \hfill
     \begin{subfigure}[b]{0.4\textwidth}
         \centering
         \includegraphics[width=\textwidth]{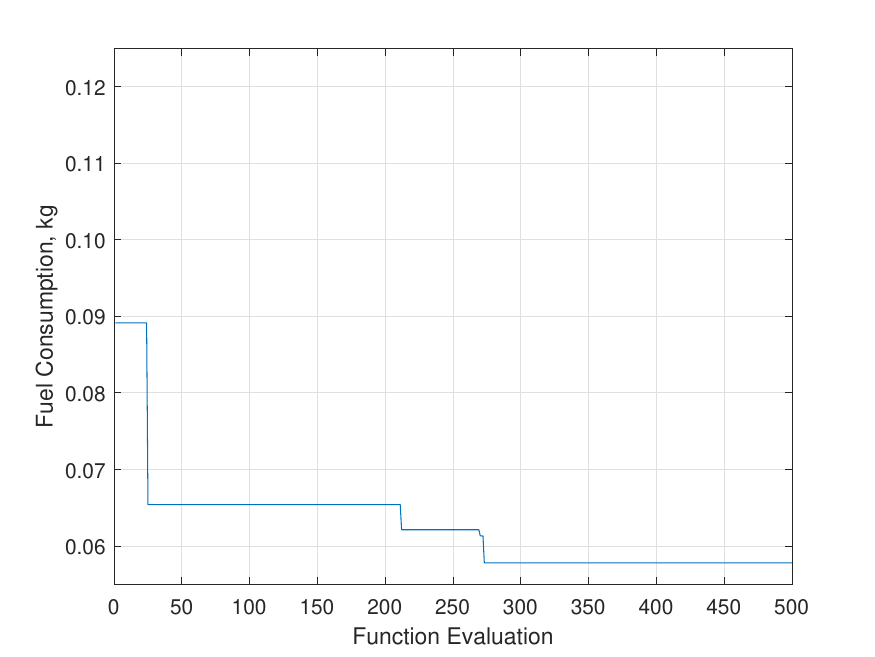}
         \caption{Case 3}
         \label{fig:opt_case3}
     \end{subfigure}
        \caption{Bayesian Optimization Minimum Objective Plot}
        \label{fig:opt_results}
\end{figure}

\begin{table*}[h]
	\begin{center}
	\caption{\label{tab:case2_3} Case Study 2 and 3 Optimal Action Variables}
		\begin{tabular}{|l|l|l|l|l|l|l|l|}
			\hline
			Methods & $\psi_{Thrust_1}$, deg & $\psi_{Thrust_2}$, deg & $\psi_{Thrust_3}$, deg & $\psi_{Thrust_4}$, deg & $\theta_{Thrust}$, deg & $m_0$, kg & $F_{Thrust}$, N\\
			\hline
            Case Study 2 & 78.4  &  37.1  &  84.3  &  40.3  &  41.7  &  2.12 & -\\
            \hline
            Case Study 3 & 78.8  &  40.9  &  87.9  &  38.2  &  40.8  &  2.07 & 5.12\\
            \hline
		\end{tabular}
	\end{center}
\end{table*}

\begin{table*}[h]
	\begin{center}
	\caption{\label{tab:fuel} Optimization and RL results comparison}
		\begin{tabular}{|l|l|l|l|}
			\hline
			Methods & Fuel Consumption, kg & Training Time Cost & Execution Time\\
            \hline
            Optimization Case 1 & 0.079 & 17 hr & 17 hr\\
			\hline
            Optimization Case 2 & 0.077 & 24.1 hr & 24.1 hr\\
            \hline
            Optimization Case 3 & 0.058 & 38.9 hr & 38.9 hr\\
            \hline
            RL & 0.083 & 511 hr & 41 ms\\
            \hline
		\end{tabular}
	\end{center}
\end{table*}

\section{Conclusion}
A semi-decentralized tether-net system was introduced with four maneuverable corner nodes (individually propelled) to control the trajectory of the net for increased robustness in capturing space debris, here represented by the second stage of the Zenit-2 launch vehicle. A reinforcement learning (RL) based approach is proposed to shift the control system design cost to a heavy but offline computation process, leading to fast-to-execute trained controllers that can be used online to control the net trajectory across various scenarios. Here scenarios are defined in terms of lateral offset between the chaser spacecraft launching the net and the target debris. The performance of the RL-based trajectory controllers is compared with optimal trajectory plans resulting from Bayesian Optimization applied to specific scenarios. The optimization-based solutions are developed for three different case study settings with increasing complexity (and increasing control authority allowed by the setting), going from simply controlling the thrust angle to also controlling the thrust force magnitude and initial fuel mass. In Case Study 3, the optimization's thrust angles, initial mass, and thrust magnitude were action variables. As expected, the third case study with the highest complexity achieves the lowest fuel consumption in the selected capture scenario, which also demonstrates the effectiveness of the implemented optimization process. 

The policy model generated by \ac{RL} is observed to perform successful capture in the same scenario as used by the optimization scenario while providing a higher success rate when the lateral shift is introduced. This demonstrates the potential generalizability of the RL-based control policy, which is intractable to achieve with optimization since a separate expansive optimization has to be run for every offset scenario with the latter. To put this into perspective, the trained RL policy executes in 50 milliseconds versus 17 hours required by optimization for a given scenario. An immediate next step in this research is to extend the RL approach to produce the control policies for Case Study 2 and 3, involving increased control authority. Further future work would entail adding more features to the scenarios over which RL is tasked to generalize, including the debris's rotational rates and sensing uncertainties during flight.

\section*{Acknowledgements}
The authors would like to thank CM Labs Simulations for providing licenses for the Vortex Studio simulation framework. This work is supported under CMMI Award numbered 2128578 from the National Science Foundation (NSF). The author's opinions, findings, and conclusions or recommendations expressed in this material do not necessarily reflect the views of the National Science Foundation.

\bibliographystyle{IEEEtran}
\bibliography{sample} 


\end{document}